\documentclass[12pt]{article}
\usepackage{amsmath}
\usepackage{graphicx,psfrag,epsf}
\usepackage{enumerate}
\usepackage{natbib}
\usepackage{url} 
\usepackage{xcolor}
\usepackage{float}
\newcommand{\bm}{\boldsymbol}
\usepackage[]{threeparttable}
\DeclareMathOperator*{\argmax}{arg\,max}

\newcommand\fnote[1]{\captionsetup{font=footnotesize}\caption*{#1}}
\usepackage{caption,subcaption,tabularx,booktabs}

\makeatletter
\let\TPT@hookin\@gobble
\let\TPT@hookarg\@gobble
\makeatother

\newcommand{\blind}{0}

\begin{document}

\def\spacingset#1{\renewcommand{\baselinestretch}%
{#1}\small\normalsize} \spacingset{1}


\if0\blind
{
  \title{\bf Fast variational Bayes methods for multinomial probit models}
  \author{Rub\'en Loaiza-Maya\\
    and \\
    Didier Nibbering\thanks{Correspondence to: Department of Econometrics \& Business Statistics, Monash University, Clayton VIC 3800, Australia, e-mail: \textsf{didier.nibbering@monash.edu}} \\
    Department of Econometrics and Business Statistics, Monash University}
  \maketitle
} \fi

\if1\blind
{
  \bigskip
  \bigskip
  \bigskip
  \begin{center}
    {\LARGE\bf Fast variational Bayes methods for multinomial probit models}
\end{center}
  \medskip
} \fi

\bigskip
\begin{abstract}
The multinomial probit model is often used to analyze choice behaviour. However, estimation with existing Markov chain Monte Carlo (MCMC) methods is computationally costly, which limits its applicability to large choice data sets. This paper proposes a variational Bayes method that is accurate and fast, even when a large number of choice alternatives and observations are considered. Variational methods usually require an analytical expression for the unnormalized posterior density and an adequate choice of variational family. Both are challenging to specify in a multinomial probit, which has a posterior that requires identifying restrictions and is augmented with a large set of latent utilities. We employ a spherical transformation on the covariance matrix of the latent utilities to construct an unnormalized augmented posterior that identifies the parameters, and use the conditional posterior of the latent utilities as part of the variational family. The proposed method is faster than MCMC, and can be made scalable to both a large number of choice alternatives and a large number of observations. The accuracy and scalability of our method is illustrated in numerical experiments and real purchase data with one million observations.
\end{abstract}

\noindent%
\vfill

\newpage
\spacingset{1.8} 

\section{Introduction}

The multinomial probit (MNP) model is a popular tool for analyzing choice behavior, with recent applications including brand choices \citep{miyazaki2021dynamic}, employment choices \citep{mishkin2021gender}, and car parking choices \citep{paleti2018generalized}. The main advantage of the MNP model is the relaxation of the independence of irrelevant alternatives assumption made by multinomial logit models. The MNP achieves this by specifying the conditional covariance matrix of the latent utilities of the choice alternatives. However, estimation of the MNP model is computationally costly. The evaluation of the likelihood function involves high-dimensional integrals, which can be solved by simulation methods.
In this paper we propose a variational Bayes (VB) method for estimation in the MNP model, which is accurate and fast even when applied to choice sets that have a large number of choice alternatives and a large number of observations. 


Bayesian analysis of the MNP model has reduced the computational complexity of parameter estimation, but its applicability to modern choice data sets is still limited.
Bayesian estimation of the MNP model augments the likelihood function with a set of latent utilities, which are then generated inside a Markov chain Monte Carlo (MCMC) scheme \citep{albert1993bayesian}. This approach avoids the computationally costly step of directly calculating the choice probabilities in the likelihood function via numerical integration. Instead, each MCMC iteration draws a vector of latent utilities from a truncated normal distribution for each observation \citep{mcculloch1994exact}. Since these draws are highly auto-correlated, a large amount of iterations are required to achieve convergence. Therefore, this approach still has a substantial computational burden, especially when the number of observations or the number of choice alternatives is large. 
  
VB is a computationally scalable alternative to MCMC. Instead of sampling from the posterior, VB calibrates a parametric approximating density by minimizing  a divergence function to the posterior. Applying VB to the MNP model poses two main challenges.
First, parameter identification in the MNP model requires a restriction on the covariance matrix \citep{bunch1991estimability}. \citet{burgette2012trace} show that a trace restriction has the best performance, and introduce a ``working parameter'' that rescales the trace within the MCMC algorithm. 
This parameter is not part of the model specification and not identified given the data. Hence an analytical expression for the unnormalized posterior density, which is required for the implementation of VB, is not available.
Second, to reduce its computational complexity, the posterior density of the MNP model has to be augmented with a large number of latent variables. Existing VB methods often make strong assumptions on the approximating densities for the latent variables \citep{westling2019beyond}.

Because of these challenges, existing VB approaches are designed for restrictive specifications of the MNP model. For instance, \citet{girolami2006variational} model the latent utilities as independent Gaussian processes, that only allow for correlations that are a function of  regressors. \citet{fasano2022class} conduct VB for the coefficients conditional on a fixed covariance matrix. Moreover, both papers impose strong independence assumptions on the family of variational approximations to the posterior density. 

This paper proposes a VB method for the MNP model that overcomes the two challenges. 
First, we construct an analytical expression of the unnormalized augmented posterior density by using the model specification as proposed by \citet{loaiza2021scalable}. They transform the covariance parameters into a spherical coordinate system. The spherical transformation naturally imposes the trace restriction on the covariance matrix, and therefore the parameters are identified within the model specification.  
Second, we use the accurate variational approximation for models with multiple latent variables proposed in \citet{loaiza2021fast}. The approximation for the latent utilities is the exact conditional posterior distribution for the latent utilities, and the approximation for the coefficients and the parameters in the spherical transformation is Gaussian.
The combination of these two innovations results in an accurate VB approach that is substantially faster than MCMC. 

Additionally, we demonstrate that our VB approach is scalable to data sets with a large number of observations.
The VB optimization problem is solved with stochastic gradient ascent (SGA), where each iteration takes a draw from the conditional posterior distribution for the latent utilities. Sampling from the conditional posterior distribution of the latent utilities is computationally costly and hence takes the majority of the computation time in VB. Since SGA allows for subsampling, which means that in each iteration only a subsample of the latent utilities have to be generated, the computational complexity of the proposed methods can be further reduced.

We formulate our VB approach for the general multivariate multinomial probit (MVMNP) model.
This means that we provide one common method for fast and accurate inference for a variety of different models, such as the MNP and multivariate probit (MVP) models, that are currently estimated with different identification strategies and different MCMC methods.
%
For instance, \citet{zhang2006sampling} and \citet{talhouk2012efficient} fix the covariance matrix to be a correlation matrix in an MVP model. 
\citet{chib1998mcmc} and \citet{mcculloch2000bayesian} fix one element of the covariance matrix in MNP models, and \citet{zhang2008bayesian} extend this to the MVMNP model. \citet{burgette2012trace} introduce the trace restriction in the MNP model, and \citet{richard2012sparse} fix the scale of a factor structure in the covariance matrix in the MVP model.  
Moreover, our method extends the spherical transformation on the covariance parameters to any MVMNP model, which allows for a factor structure in the covariance matrix that can substantially reduce the number of parameters to be estimated. Hence, our method can be applied to data sets with a large number of choices and a large number of choice alternatives.

Numerical experiments show that our VB method provides accurate parameter estimates and choice probabilities, while it only takes a fraction of the computational cost of MCMC.
A numerical experiment with a small data set of 10,000 observations, in which MCMC is feasible, shows a minimal loss in predictive accuracy of VB relative to MCMC. An experiment with one million observations, in which MCMC is infeasible, shows that VB applied to a large data set can improve predictive accuracy relative to MCMC applied to only a subset of the observations. 

We illustrate the practical relevance of our method with two empirical applications.
First, VB produces similar results as MCMC around 10\% of the computation time, in a small real data set of laundry detergent purchases.
The second application considers a large-scale choice data set.
Data on large choice sets with a large number of observations is widely available nowadays. We estimate the MNP model with the proposed VB method on more than a million pasta purchases in less than 1.5 hours, while MCMC takes more than 90 hours.


The outline of the remainder of this paper is as follows. Section~\ref{sec:spec} discusses the model specification and Section~\ref{sec:estimation} develops our VB method. Section~\ref{sec:numerical} conducts numerical experiments to evaluate its accuracy and computational costs, and Section~\ref{sec:application} applies the proposed methods to real choice data sets. Section~\ref{sec: conclusion} concludes.

\section{Model specification}\label{sec:spec}

\subsection{Multivariate multinomial probit model}\label{sec:mvmnp}
We observe $K$ multinomial choices, where each choice $k=1,\dots,K$ has $J_k+1$ choice alternatives, for individual $i=1,\dots,N$. Let $\bm{Y}_i=(Y_{i1},\dots,Y_{iK})^\top$ denote the $K$-dimensional random variable describing the joint set of choices for individual $i$, where $Y_{ik}=j$ if individual $i$ chooses $j=0,1,\dots,J_k$ for the $k$-th choice. The number of potential outcomes of $\bm{Y}_i$ is $\prod_{k=1}^K(J_k+1)$. 

Assume that for the $k$-th choice there is a $J_k$-dimensional vector  $\bm{Z}_{ik}=(Z_{ik1},\dots,Z_{ikJ_k})^\top$ of continuous random variables representing the latent utilities for the choice alternatives, and which excludes the base category latent utility $Z_{ik0}$.
The multinomial outcome $Y_{ik}$ is determined by the maximum value of $\bm{Z}_{ik}$ as follows:
\begin{align}\label{eq:Y_ik}
    Y_{ik} = \begin{cases} 0 & \text{ if } \max(\bm{Z}_{ik})<0,\\
    j & \text{ if } Z_{ikj}=\max(\bm{Z}_{ik})>0,\end{cases}
\end{align}
where $\max(\bm{Z}_{ik})$ is the largest element of $\bm{Z}_{ik}$. The latent utilities corresponding to the choice alternatives in choice $k$ are modeled as
\begin{align}\label{eq:Z_ik}
    \bm{Z}_{ik} = X_{ik}\bm{\beta}_k +\bm{\varepsilon}_{ik},
\end{align}
where $X_{ik}$ is a $J_k \times r_k$ regressor matrix, $\bm{\beta}_{k}$ is an $r_k$-dimensional vector of coefficients and $\bm{\varepsilon}_{ik}=(\varepsilon_{ik1},\dots,\varepsilon_{ikJ_k})^\top$ is a $J_k$-dimensional disturbance vector with mean zero.

The regressor matrix $X_{ik}$ typically includes $J_k$ choice alternative-specific intercepts, an $n_d$-dimensional vector $\bm{x}_{i}^d$ of individual-specific characteristics, and a $(J_k+1) \times n_a$ matrix $X_{ik}^a$ of $n_a$ alternative-specific covariates, such that $r_k = J_k+J_kn_d+n_a$ and
\begin{align}\label{eq:X_ik}
    X_{ik} = [I_{J_k} \quad (\bm{x}_{i}^d)^\top\otimes I_{J_k} \quad T_kX_{ik}^a],
\end{align}
with transformation matrix $T_k=[-\bm{\iota}_{J_k} \quad I_{J_k}]$,  
where $\bm{\iota}_{J_k}$ denotes the $J_k$-dimensional vector of ones and $I_{J_k}$ the $J_k\times J_k$ identity matrix.

The model specified in \eqref{eq:Y_ik} and \eqref{eq:Z_ik} only considers the $J_k$ utilities in $\bm{Z}_{ik}$. Therefore,  the covariates in the regressor matrix $X_{ik}$ in \eqref{eq:X_ik} are transformed to match the dimensions of $\bm{Z}_{ik}$. The transformation matrix $T_k$ subtracts the covariates corresponding to choice category $j=0$ in choice $k$ from the covariates corresponding to the remaining choice alternatives in choice $k$. This model specification addresses the first parameter identification problem in the MVMNP model: additive redundancy arises if a unique utility is specified for each choice alternative, as discussed in \citet{bunch1991estimability}. The second identification problem is caused by multiplicative redundancy: multiplying both sides of \eqref{eq:Z_ik} by a positive scalar does not change $Y_{ik}$ in \eqref{eq:Y_ik}. We address this identification problem in Section~\ref{sec:identification}.

\sloppy By specifying a joint distribution for $\bm{\varepsilon}_{i} = (\bm{\varepsilon}_{i1}^\top,\dots,\bm{\varepsilon}_{iK}^\top)^\top$, the MVMNP model can allow for within-choice and between-choice correlation in the latent utilities. 
The $K$ models implied by \eqref{eq:Z_ik} can be stacked to obtain the multivariate utility model
\begin{align}
    \bm{Z}_i = X_i\bm{\beta}+\bm{\varepsilon}_i, \quad \bm{\varepsilon}_{i}  \sim N[\bm{0}_J,\Sigma],
\end{align}
where $\bm{Z}_i=(\bm{Z}_{i1}^\top,\dots,\bm{Z}_{iK}^\top)^\top$ is a $J-$dimensional vector with $J=\sum_{k=1}^K J_k$, $X_i=\text{blockdiag}(X_{i1},\dots,X_{iK})$ is a $J\times r$ block diagonal regressor matrix with $r=\sum_{k=1}^K r_k$, and $\bm{\beta}=(\bm{\beta}_1^\top,\dots,\bm{\beta}_K^\top)^\top$ is an $r$-dimensional vector of coefficients.
The $J \times J$ covariance matrix of the disturbance vector $\bm{\varepsilon}_{i}$ can be represented as 
\begin{align}\label{eq:Sigma}
    \Sigma= 
    \left[\begin{array}{cccc}
\Sigma_{11}&\Sigma_{12}&\cdots &\Sigma_{1K}\\
\Sigma_{21}&\Sigma_{22}&\cdots &\Sigma_{2K}\\
\vdots & &\ddots &\vdots \\
\Sigma_{K1}&\Sigma_{K2}&\cdots &\Sigma_{KK}\\
\end{array}\right],
\end{align}
where $\Sigma_{kl}=\Sigma_{lk}$. The $J_k \times J_l$ covariance matrix $\Sigma_{kl} = \text{cov}(\bm{\varepsilon}_{ik},\bm{\varepsilon}_{il})$ captures the correlation across the utilities within-choice $k$ if $k=l$, and the correlation across the utilities between choices $k$ and $l$ if $k \neq l$, with $k=1,\dots,K$ and $l=1,\dots,K$. 

The MVMNP model easily simplifies to other commonly used choice models. First, for $K=1$ the MVMNP boils down to a multinomial probit (MNP) model, with $J+1$ choice alternatives corresponding to potentially correlated latent utilities. When $K>1$ and the elements of $\Sigma_{lk}$ equal zero for all $k\neq l$ in \eqref{eq:Sigma}, we have $K$ independent MNP models. Second, with $J_k=1$ for $k = 1,\dots,K$, we have a multivariate probit (MVP) model, with $K$ potentially correlated binary choices. When $K=1$ and $J=1$, the model boils down to a simple binary probit model. 

For the remainder of the paper we stack the regressor matrices for all individuals in the $(NJ)\times r$ matrix $X = \left[X_1^\top |\dots| X_N^\top\right]^\top$ , the random choice vectors in $\bm{Y} = \left(\bm{Y}_1^\top,\dots,\bm{Y}_N^\top\right)^\top$, and the random latent utility vectors in $\bm{Z} = \left(\bm{Z}_1^\top,\dots,\bm{Z}_N^\top\right)^\top$.

\subsection{Factor structure covariance matrix}\label{sec:factor}
The total number of unique parameters in the covariance matrix $\Sigma$ equals $J(J+1)/2$, which grows quadratically in the number of choices and the number of choice alternatives. Since these parameters have to be estimated from a single multinomial variable $Y_i$, accurate parameter estimation is challenging if either $J$ or $K$ is large, or both, even when a relatively large number of observations $N$ is available. 

To reduce the dimension of the parameter space, we specify a factor structure for $\Sigma$. Define the $J\times p$ matrix $B$ with $p \leq J$ and the $J\times J$ diagonal matrix $D$. We model $\Sigma$ as 
\begin{align}\label{eq:factor}
    \Sigma = BB^\top+D^2.
\end{align}
The total number of parameters in $B$ and $D$ is $n=J(p+1)$. This implies that for a given value of $p$, the number of parameters grows linearly with $J$, instead of quadratically.

To understand the implications of this factor structure on the within- and between-choice correlations, the matrices $B$ and $D$ are partitioned as $B = \left[B_1^\top |\dots| B_K^\top\right]^\top$ and $D = \text{diag}\left(\bm{d}_1^\top,\dots,\bm{d}_K^\top\right)$, where $B_k$ is a $J_k\times p$ matrix and $\bm{d}_k$ a $J_k$-dimensional vector corresponding to choice $k$. The within-choice covariance matrix $\Sigma_{kk}$ is expressed as
\begin{align}\label{eq:Sigma_kk}
    \Sigma_{kk} = B_kB_k^\top+D_k^2,
\end{align}
which shows that the matrices $\{\Sigma_{kk}\}_{k=1}^K$, and hence the within-choice correlations for each choice, are characterised by disjoint sets of model parameters. The between-choice covariance matrix $\Sigma_{kl}$ with $k \neq l$ is given by
\begin{align}
    \Sigma_{kl}=B_kB_l^\top,
\end{align}
which is only a function of the matrices $B_k$ and $B_l$ corresponding to the within-choice correlations in choices $k$ and $l$. In sum, the matrices $\{B_{k}\}_{k=1}^K$ determine both the within- and between-choice covariances in the latent utilities. 

A factor structure with a small number of factors $p$ is correctly specified if the variability in the $J$ latent utilities in the data generating process can be captured by the $p$ latent factors. This is the case if, for instance, 
choice behavior is driven by a small number of unobserved features of the choice alternatives. 
Another example is the MVMNP model with uncorrelated choices, which can be estimated as separate univariate MNP models in which the factors only need to capture within-choice correlations.
Issues of model misspecification may arise when the total number of unknown underlying factors is greater than $p$.
This may be the case if choice behavior is driven by a large number of underlying choice features, or if the correlation pattern 
cannot be captured by a few factors. For instance, if choice alternatives are spatially related, the covariance matrix may have a banded pattern.

Even when the covariance matrix is misspecified, a low dimensional factor structure could potentially be favoured to trade-off flexibility for parsimony in the model, provided that key outputs from the model, such as predictive performance, remain accurate.
The factor analysis literature has proposed several approaches to selecting $p$ optimally, including the use of information criteria, marginal likelihoods and cross-validation \citep{fruhwirth2018sparse}, and these approaches may also be applied to the MVMNP model.
We set $p=K$ in this paper, which allows for different covariance structures within each choice while reducing the number of covariance parameters to be estimated, and hence reducing parameter uncertainty and computation time.

\subsection{Parameter identification}\label{sec:identification}
As discussed in Section~\ref{sec:mvmnp}, the scale of the latent utilities is unidentified. 
To identify the parameters, we extend the approach of Loaiza-Maya and Nibbering (2021) from an MNP model to the MVMNP model. For each choice $k$, we fix the scale using  $\text{trace}(\Sigma_{kk})=J_k$, by transforming the elements of $B_k$ and $\bm{d}_k$ into a spherical coordinate system.

This parameter identification strategy has three advantages. 
First, in contrast to alternative identification restrictions, the trace restriction identifies the model parameters without fixing specific elements in the covariance matrix $\Sigma$. \citet{burgette2012trace} show that Bayesian estimation in the MNP model is sensitive to which elements in the covariance matrix are fixed. 
Second, the spherical transformation on the factor covariance structure simplifies parameter estimation, as it naturally satisfies the trace restriction. Instead of performing inference on a parameter space with a joint restriction on all elements in each $\Sigma_{kk}$, we perform inference on the angle parameter space for which no joint parameter restriction is required.
Third, since the spherical transformation naturally imposes the trace restriction, our approach does not require rescaling of the covariance matrix. Therefore, an analytical expression for the gradient of the likelihood function is available, which allows us to apply VB. Even with $p=J$, in which case there is no dimension reduction, writing $\Sigma$ as in \eqref{eq:factor} has the benefit that it can be transformed by a spherical transformation and VB can be applied. 

The spherical transformation is applied to the vector $\bm{\psi}_k$, which is constructed from the elements of $B_k$ and $\bm{d}_k$ as
\begin{align}\label{eq: psi}
    \bm{\psi}_k = \left(\psi_{k1},\dots,\psi_{kn_k}\right)^\top = (\text{vec}(B_k)^\top,\bm{d}_k^\top)^\top,
\end{align}
where $n_k=J_k(p+1)$, $\text{vec}(\cdot)$ denotes the vectorization operator, and $\text{trace}(\Sigma_{kk})=\sum_{l=1}^{n_k}\psi_{kl}^2$. We transform $\bm{\psi}_{k}$ into a spherical coordinate system that is defined by a radius, which we set to $\sqrt{J_k}$, and an $(n_k-1)$-dimensional vector of angles $\bm{\kappa}_k = \left(\kappa_{k1},\dots,\kappa_{k,n_k-1}\right)^\top$.

The spherical transformation reparameterises $\bm{\psi}_k$ in terms of $\bm{\kappa}_k$ as
\begin{equation}\label{eq:transformation}
  \psi_{kl}(\bm{\kappa}_k) = \begin{cases}
      \sqrt{J_k}\cos\kappa_{k1} & \text{for $l=1$},\\
    \sqrt{J_k}\cos\kappa_{kl}\prod_{j = 1}^{l-1}\sin\kappa_{kj} &   \text{for $1<l<n_k$},\\
    \sqrt{J_k}\prod_{j = 1}^{l-1}\sin\kappa_{kj} & \text{for $l=n_k$},
  \end{cases}
\end{equation}
where $\kappa_{kl} \in [0,\pi)$ for $l<n_k-J_k+1$. The remaining angle bounds, $\kappa_{kl} \in [0,\frac{\pi}{2})$ for $n_k-J_k+1\le l \le n_k-1$, ensure that the elements of $\bm{d}_k$ are strictly greater than zero. This ensures that the map from $\bm{d}_k$ to $\Sigma_{kk}$ in \eqref{eq:Sigma_kk} is bijective.

The transformation in \eqref{eq:transformation} satisfies $\sum_{l=1}^{n_k} \psi_{kl}({\bm{\kappa}_k})^2 = J_k$ for any value of $\bm{\kappa}_k$. This reparametrization is applied to all $\Sigma_{kk}$, which results in a covariance matrix $\Sigma$ that is characterised by the $n$-dimensional vector $\bm{\kappa} = \left(\bm{\kappa}_1^\top,\dots,\bm{\kappa}_K^\top\right)^\top$, where $n=\sum_{k=1}^Kn_k$. The vector $\bm{\kappa}$ imposes $K$ trace restrictions simultaneously: one for each choice.

The inverse function of the spherical transformation in \eqref{eq:transformation} is
\begin{equation}\label{eq:inverse_transformation}
  \kappa_{kl}(\bm{\psi}_k) = \begin{cases}
      \arccos\left[\psi_{kl}\left(\sum_{j=l}^{n_k}\psi_{kj}^2\right)^{-\frac{1}{2}}\right] & \text{for $l<n_k-1$},\\
      \arccos\left[\psi_{kl}\left(\sum_{j=l}^{n_k}\psi_{kj}^2\right)^{-\frac{1}{2}}\right] &   \text{for $\{l =n_k-1 \land \psi_{kn_k}\ge0\}$},\\
    2\pi-\arccos\left[\psi_{kl}\left(\sum_{j=l}^{n_k}\psi_{kj}^2\right)^{-\frac{1}{2}}\right] & \text{for $\{l =n_k-1 \land \psi_{kn_k}<0\}$},
  \end{cases}
\end{equation} 
where $\kappa_{kl}=0$ if $\psi_{kl}>0$ and $\psi_{k,l+1}=\dots=\psi_{kn_k}=0$, and $\kappa_{kl}=\pi$ if $\psi_{kl}<0$ and $\psi_{k,l+1}=\dots=\psi_{kn_k}=0$.

\section{Bayesian estimation}\label{sec:estimation}
This section develops a Bayesian method for estimating the parameters in the $m$-dimensional vector $\bm{\theta} = \left(\bm{\beta}^\top,\bm{\kappa}^\top\right)^\top$, with $m = r+n$. We conduct inference of the augmented posterior density
\begin{equation}
     p(\bm{\theta},\bm{z}|\bm{y},X)\propto p(\bm{y},\bm{z}|X,\bm{\theta}) p(\bm{\theta}),\label{Eq:exact_posterior}
\end{equation}
where we use lower case letters to denote realised values of the corresponding random vectors. For instance, $\bm{y}$ is the realised vector of $\bm{Y}$.  

The augmented likelihood function is given by
\begin{align}
  p(\bm{y},\bm{z}|X,\bm{\theta}) & = p(\bm{y}|\bm{z}) p(\bm{z}|X,\bm{\theta})= \prod_{i=1}^{N}p(\bm{y}_i|\bm{z}_i) \phi_{J}\left(\bm{z}_i;X_i\bm{\beta},\Sigma(\bm{\kappa})\right),
\end{align}
where $\phi_{J}\left(\bm{z}_i;X_i\bm{\beta},\Sigma(\bm{\kappa})\right)$ denotes the $J$-variate normal density with mean $X_i\bm{\beta}$ and covariance matrix $\Sigma(\bm{\kappa})$, with $\Sigma(\bm{\kappa})$ the covariance matrix constructed from the vector of angles $\bm{\kappa}$, $p(\bm{y}_i|\bm{z}_i) = \prod_{k=1}^{K}p(y_{ik}|\bm{z}_{ik})$ and 
\begin{align}\label{eq:y_ik}
    p(y_{ik}|\bm{z}_{ik}) = \begin{cases} I\left[z_{iky_{ik}}=\text{max}(\bm{z}_{ik})\right]& \text{ if } \text{max}(\bm{z}_{ik})>0,\\
    I(y_{ik}=0) & \text{ if } \text{max}(\bm{z}_{ik})\le0 ,\end{cases}
\end{align}
where $I[A]$ is an indicator function that equals one if $A$ is true and zero otherwise.

We set the prior density as $p(\bm{\theta}) = p(\bm\beta)\prod_{k=1}^{K}p(\bm\kappa_k)$, with $\bm{\beta}\sim N(\bm{0}_r, \frac{1}{10}I_r)$ and $p(\bm\kappa_k)$ specified in online appendix~A, with an implied prior mean for $\Sigma$ that equals the equicorrelated covariance matrix $\frac{1}{2}(I_J + \iota_J \iota_J^\top)$.

\subsection{Markov Chain Monte Carlo sampling}
The posterior density in \eqref{Eq:exact_posterior} can be computed using MCMC sampling. For each individual, the latent utility of each choice alternative is sampled conditional on all the other choice alternatives from a truncated normal. This process induces a sequence of latent utility draws that is highly auto-correlated. Therefore MCMC requires a large number of iterations such that convergence is achieved. Since each iteration involves $N \times J$ draws from a truncated normal, MCMC is computationally costly. Online appendix~B describes the MCMC sampling scheme for the MVMNP model.

The computational costs of an MCMC sampling scheme increase in the number of choice alternatives $J_k$ in each choice $k$, the number of choices $K$, and the number of observations $N$. As a result, when the total number of choice alternatives $J$ is large, MCMC is considered to be computationally practical as long as the number of observations is small. There are two empirical settings where this is the case. First, univariate choice sets ($K=1$) that have a large number of choice alternatives. Second, applications that consider multiple choices and for which the overall number of choice alternatives $J$ is large.
However, it is precisely in these type of settings where having a large number of observations is key for accurate estimation of the high-dimensional covariance matrix of the latent utilities.

\subsection{Variational Bayes}
To circumvent the computational challenges of MCMC, we utilize variational Bayes. 
VB approximates the posterior density in \eqref{Eq:exact_posterior} by a parametric density $q_{\widehat{\lambda}}\left(\bm{\theta},\bm{z}\right)\in\mathcal{Q}$ from the class of density functions $\mathcal{Q} = \{q_\lambda\left(\bm{\theta},\bm{z}\right): \bm{\lambda}\in\Lambda\}$, where $q_\lambda\left(\bm{\theta},\bm{z}\right)$ is indexed by the variational parameter vector $\bm{\lambda}\in\Lambda$. 
The optimal variational parameter $\widehat{\bm{\lambda}}$ is obtained by maximizing the evidence lower bound (ELBO) function $\mathcal{L}\left(\bm{\lambda}\right)=E_{q_{\lambda}}\left[\log g(\bm{\theta},\bm{z})-\log q_\lambda(\bm{\theta},\bm{z})\right]$:
\begin{align}
    \widehat{\bm{\lambda}}&= \argmax_{\bm{\lambda}\in\Lambda}E_{q_{\lambda}}\left[\log g(\bm{\theta},\bm{z})-\log q_\lambda(\bm{\theta},\bm{z})\right],\label{Eq:optimization}
\end{align}
where $g(\bm{\theta},\bm{z}) = p(\bm{y}|\bm{z})p(\bm{z}|X,\bm{\theta})p(\bm{\theta})$ is the unnormalized posterior density. \citet{ormerod2010explaining} show that the optimization problem in \eqref{Eq:optimization} is equivalent, but computationally more efficient, to minimizing the Kullback-Leibler (KL) divergence between $q_{{\lambda}}\left(\bm{\theta},\bm{z}\right)$ and the exact posterior density $p(\bm{\theta},\bm{z}|\bm{y},X)$.

While MCMC generates draws from the exact posterior distribution, VB can only construct an approximation to it. However, VB has three main advantages over MCMC for the MVMNP model.
First, MCMC may show high auto-correlation in its chain for this model, leading to substantial computational costs. VB relies on optimization rather than sampling, and therefore reduces the computation time.
Second, VB requires much less storage memory as the output from VB is the calibrated parameter vector of the approximation, rather than a large number of parameter draws. Third, VB can readily incorporate subsampling of the latent utilities in the optimization routine, which can further reduce the computational burden.

\subsubsection{Variational family}\label{sec:vf}
Key to the implementation of VB is the choice of the variational family $\mathcal{Q}$.
We set $q_\lambda\left(\bm{\theta},\bm{z}\right) = p(\bm{z}|\bm{\theta},\bm{y},X)q_\lambda(\bm{\theta})$, where $p(\bm{z}|\bm{\theta},\bm{y},X)$ is the conditional posterior of the latent utilities defined in online appendix~B, and we define $q_\lambda(\bm{\theta})$ below. 
\cite{loaiza2021fast} show that the optimization problem in \eqref{Eq:optimization} with this variational family is equivalent to the optimization problem that considers the KL divergence between the intractable posterior $p(\bm{\theta}|\bm{y},X)$ and $q_\lambda(\bm{\theta})$.
Alternative specifications for the variational family may result in approximating errors for the latent utilities, which can provide inconsistent estimates, as shown in \cite{westling2019beyond}.

For the choice of $q_\lambda(\bm{\theta})$, we follow \cite{ong2018gaussian} and employ a Gaussian density with mean $\bm{\mu}$ and covariance matrix $\Omega = CC^\top+E^2$, where $C$ is a matrix of dimension $m\times s$ for $s<m$, $E=\text{diag}(\bm{e})$ and $\bm{e}$ an $m$-dimensional vector. The variational parameter vector for this approximating class is $\bm{\lambda} = \left(\bm{\mu}^\top,\text{vech}(C)^\top,\bm{e}^\top\right)^\top$, where the operator vech denotes the half vectorization of a rectangular matrix  such that $\text{vech}(C)= \left(C_{1:m,1}^\top,\dots,C_{s:m,s}^\top\right)^\top$ with $C_{j:m,j} = \left(C_{jj},\dots,C_{mj}\right)^\top$ for $j = 1,\dots,s$. 

\subsubsection{Stochastic gradient ascent}
We solve the optimization problem in \eqref{Eq:optimization} using SGA methods. SGA calibrates the variational parameter by iterating over
\begin{equation}\label{eq: lambda_iter}
    \bm{\lambda}^{[j+1]} = \bm{\lambda}^{[j]} + \bm{\rho}^{[j]}\circ\widehat{\nabla_\lambda \mathcal{L}\left(\bm{\lambda}^{[j]}\right)}, 
\end{equation}
until convergence is achieved. The vector $\bm{\rho}^{[j]}$ contains the so called ``learning parameters'', which we set according to the ADADELTA approach in \cite{zeiler2012adadelta}. The vector $\widehat{\nabla_\lambda \mathcal{L}\left(\bm{\lambda}^{[j]}\right)}$ is an unbiased estimate of the gradient of the ELBO 
evaluated at $\bm{\lambda}^{[j]}$. 

We construct $\widehat{\nabla_\lambda \mathcal{L}\left(\bm{\lambda}\right)}$ using the following expression of the gradient
\begin{equation}\label{Eq:gradient}
\nabla_\lambda \mathcal{L}\left(\bm{\lambda}\right) =E_{\bm{z},\bm{\zeta}}\left[\frac{\partial\bm{\theta}(\bm{\zeta},\bm{\lambda}) }{\partial\bm{\lambda}}^\top\left\{{\nabla_{\theta}\log g}\left[\bm{\theta}(\bm{\zeta},\bm{\lambda}),\bm{z}\right]-\nabla_\theta \log q_\lambda\left[\bm{\theta}(\bm{\zeta},\bm{\lambda}) \right]\right\}\right],
\end{equation}
where $\bm{\theta}(\bm{\zeta},\bm{\lambda}) = \bm{\mu}+C\bm{w}+\bm{e}\circ\bm{\epsilon}$, ${\bm{w}}\sim N(\bm{0}_{s},I_{s})$, $\bm{\epsilon}\sim N(\bm{0}_{m},I_{m})$, and
$\bm{\zeta} = \left({\bm{w}}^\top,{\bm{\epsilon}}^\top\right)^\top$. 
This expression is derived in \cite{loaiza2021fast} using the ``re-parametrization trick'' in \citet{kingma2013auto}. 
We derive $\nabla_{\theta}\log g\left[\bm{\theta}(\bm{\zeta},\bm{\lambda}),\bm{z}\right]$ for the MVMNP model in online appendix~C, and $\frac{\partial\bm{\theta}(\bm{\zeta},\bm{\lambda}) }{\partial\bm{\lambda}}$ and $\nabla_\theta \log q_\lambda\left[\bm{\theta}(\bm{\zeta},\bm{\lambda}) \right]$ are provided in \cite{ong2018gaussian}.

An unbiased estimate of $\nabla_\lambda \mathcal{L}\left(\bm{\lambda}\right)$ is constructed by using a sample estimate of the expectation. At each SGA iteration $[j]$, we calculate a sample estimate of \eqref{Eq:gradient} based on only one draw for both $\bm\zeta$ and $\bm{z}$:  $\bm\zeta^{[j]}\sim N(\bm{0}_{s+m},I_{s+m})$ and  $\bm{z}^{[j]}\sim p(\bm{z}|\bm{\theta}(\bm{\zeta}^{[j]},\bm{\lambda}^{[j]}),\bm{y},X)$.
Since sampling directly from $p(\bm{z}|\bm{\theta}(\bm{\zeta},\bm{\lambda}),\bm{y},X)$ is infeasible, we generate the latent utility vector draw $\bm{z}^{[j]}$ via $G$ Gibbs steps of the truncated normal algorithm proposed in \cite{mcculloch1994exact}. The Gibbs sampling algorithm is started at the last iterate value $\bm{z}^{[j-1]}$.
Since these Gibbs draws
are highly correlated, a larger value for $G$ increases the accuracy of the estimate for the gradient. However, each Gibbs step includes $N\times J$ draws from a truncated normal distribution, which is computationally costly. We find that $G=10$ balances well accuracy and computational speed . 

\subsubsection{Subsampling of the latent utilities}

The objective of VB is to compute the optimal variational parameter vector, so it suffices to run enough SGA iterations until convergence is reached for all the elements of $\bm{\lambda}^{[j]}$.
SGA generally requires a small number of iterations, making it much faster than MCMC. However, the majority of the computation time is still spent on the generation of the latent utilities. 
Since SGA allows for subsampling of the observations, the computational burden of the latent utilities can be substantially reduced in VB.

Instead of sampling the latent utilities $\bm{z}_i$ at each iteration for all individuals, SGA can estimate the gradient unbiasedly using only a subsample of the latent utilities.  
The ELBO gradient can be rewritten in terms of the variable $A\subset\{1,\dots,N\}\sim f(A)$, where a draw from $f(A)$ is a random subsample of indexes without replacement. Define the subsample of latent utilities as $\bm{z}_A = \{\bm{z}_i\}_{i\in A}$. Since  $E_A\left[{\nabla_{\theta}\log g}(\bm{\theta},\bm{z}_A)\right]={\nabla_{\theta}\log g}(\bm{\theta},\bm{z})$, it holds that
\begin{equation}\label{Eq:gradient2}
\nabla_\lambda \mathcal{L}\left(\bm{\lambda}\right) =E_{\bm{z},\bm{\zeta},A}\left[\frac{\partial\bm{\theta}(\bm{\zeta},\bm{\lambda}) }{\partial\bm{\lambda}}^\top\left\{{\nabla_{\theta}\log g}(\bm{\theta}(\bm{\zeta},\bm{\lambda}),\bm{z}_A)-\nabla_\theta \log q_\lambda\left[\bm{\theta}(\bm{\zeta},\bm{\lambda}) \right]\right\}\right].
\end{equation}
Online appendix C provides the expression for ${\nabla_{\theta}\log g}(\bm{\theta},\bm{z}_A)$ required to compute the subsampling gradient estimate for the MNP model.
An unbiased estimate of \eqref{Eq:gradient2} is constructed using a sample estimate of the expectation, using the draws $A^{[j]}\sim f(A)$, $\bm{\zeta}^{[j]}\sim N(\bm{0}_{s+m},I_{s+m})$, and $\bm{z}_{A^{[j]}}^{[j]}\sim p(\bm{z}_{A^{[j]}}|\bm{\theta}(\bm{\zeta}^{[j]},\bm{\lambda}^{[j]}),\bm{y},X)$. 

\subsubsection{Variational predictive distribution}
The VB predictive probability mass function for $\bm{Y}_{i}$ is given by
\begin{align}\label{eq:predictive}
   p_{\hat{\lambda}}(\bm{Y}_i|X_i) = \int p(\bm{Y}_i|\bm{z}_i)p(\bm{z}_i|\bm{\theta},X_i)q_{\hat{\lambda}}(\bm{\theta})d\bm{\theta}d\bm{z}_i,
\end{align}
where $X_i$ denotes the attributes of the observation $i$ to be predicted. We construct an estimate $\hat{p}_{\hat{\lambda}}(\bm{Y}_i|X_i)$ for  \eqref{eq:predictive} as the empirical probability mass implied by the draws $\{\bm{y}_i^{[m]}\}_{m=1}^M$, obtained by drawing from $\bm{\theta}^{[m]}\sim q_{\hat{\lambda}}(\bm{\theta})$, $\bm{z}_i^{[m]}\sim p(\bm{z}_i|\bm{\theta}^{[m]},X_i)$ and $\bm{y}_i^{[m]}\sim p(\bm{Y}_i|\bm{z}_i^{[m]})$.

To evaluate predictive performance, we employ the logarithmic score (log-score), which is a probabilistic measure of predictive accuracy. The choice-specific log-score is given as
\begin{align}\label{eq:logscore}
    \text{log-score}_k = \frac{1}{N}\sum_{i=1}^N&\ln(\hat{p}_{\hat{\lambda}}({Y}_{ik}|X_{ik})),
\end{align}
and the total model fit can be assessed by the average log-scores across choices.

The point forecast $\hat{\bm{Y}}_i$ for $\bm{Y}_i$ is constructed as the mode of $\hat{p}_{\hat{\lambda}}(\bm{Y}_i|X_i)$. The point prediction accuracy can be measured in terms of the hit-rate given as
\begin{align}\label{eq:hitrate}
    \text{hit-rate}_k = \frac{1}{N}\sum_{i=1}^N I[\hat{Y}_{ik}=Y_{ik}].
\end{align}
For both the hit-rate and the log-score large values are preferred.

\section{Numerical experiments}\label{sec:numerical}
This section presents two numerical experiments to assess the accuracy and the computational costs of the proposed VB approach. The first experiment compares VB to MCMC in a moderately sized data set in which MCMC is computationally feasible. The second experiment is on a large dataset for which VB estimation is feasible, but MCMC is not.

\subsection{Design}\label{sec:design}
We generate a data set from the data generating process specified in \eqref{eq:Y_ik} and \eqref{eq:Z_ik} with $K=2$ and $J_1=J_2=10$. The elements of the matrices $X_{i1}^a$ and $X_{i2}^a$ are independently generated from normal distributions with corresponding mean $\mu=0$ and variance $\sigma^2=1$. These elements can be interpreted as the logarithm of the prices of the choice categories. 
We do not include individual-specific characteristics $\bm x_{i}^d$.

The true parameter vector $\bm\beta_0$ consists of $J$ intercepts drawn independently from uniform distributions $U(-0.5,0)$, and the coefficients for $X_{i1}^a$ and $X_{i2}^a$ are fixed at -0.3 and -0.6, respectively. The true covariance matrix $\Sigma_0$ is set as a draw from the inverse Wishart distribution with equicorrelated scale matrix $\frac{1}{2}(I_J + \iota_J \iota_J^\top)$ and degrees of freedom $J+3$.

We apply our VB method to two generated data sets, one with $N=10,000$ and the second one with $N=1,000,000$. For both settings we generate an additional 10,000 observations for out-of-sample evaluation. VB with subsampling is denoted by VB($\frac{M}{N}100\%$), where $\frac{M}{N}100\%$ denotes the percentage of the total estimation sample $N$ used in each VB iteration step. For the purpose of comparison, we also estimate an MVMNP model with the covariance matrix fixed at the identity matrix (VB-I), as described in online appendix~D.

The VB methods estimate the model with 5000 iterations of SGA, with 10 Gibbs sampling steps in each SGA iteration. We take 10,000 draws from the variational posterior and predictive distribution to construct the results. The results from MCMC sampling are based on 200,000 iterations, from which the first 100,000 are discarded and we use a thinning value of 10. This results in 10,000 draws from the posterior and predictive distribution. All methods use the prior specification as discussed in Section~\ref{sec:estimation}. The methods are implemented in a HP Z240 SFF Workstation with an Intel i7-7700 CPU \@ 3.6GHz.

\subsection{Results with 10,000 observations}

\subsubsection{Convergence and computation time}
First, we assess convergence of SGA in our VB methods. The ELBO in \eqref{Eq:optimization}, which is typically used as convergence measure in VB, is not available in closed-form. The hit-rate defined in \eqref{eq:hitrate}, with $\hat{\bm{Y}}_i$ as the mode of $\hat{p}_{\lambda^{[j]}}(\bm{Y}_i|X_i)$ in iteration $[j]$, can be used instead. To reduce the computational costs, we evaluate the conditional hit-rate for a fixed random subsample of 500 observations, using 200 draws from $\hat{p}_{\lambda^{[j]}}(\bm{Y}_i|X_i)$, in each tenth iteration.  

\begin{figure}[tb!]
\caption{VB conditional hit-rate in numerical experiment with 10,000 observations}
\centering
\includegraphics*[width=\textwidth,trim = 0 0 0 0]{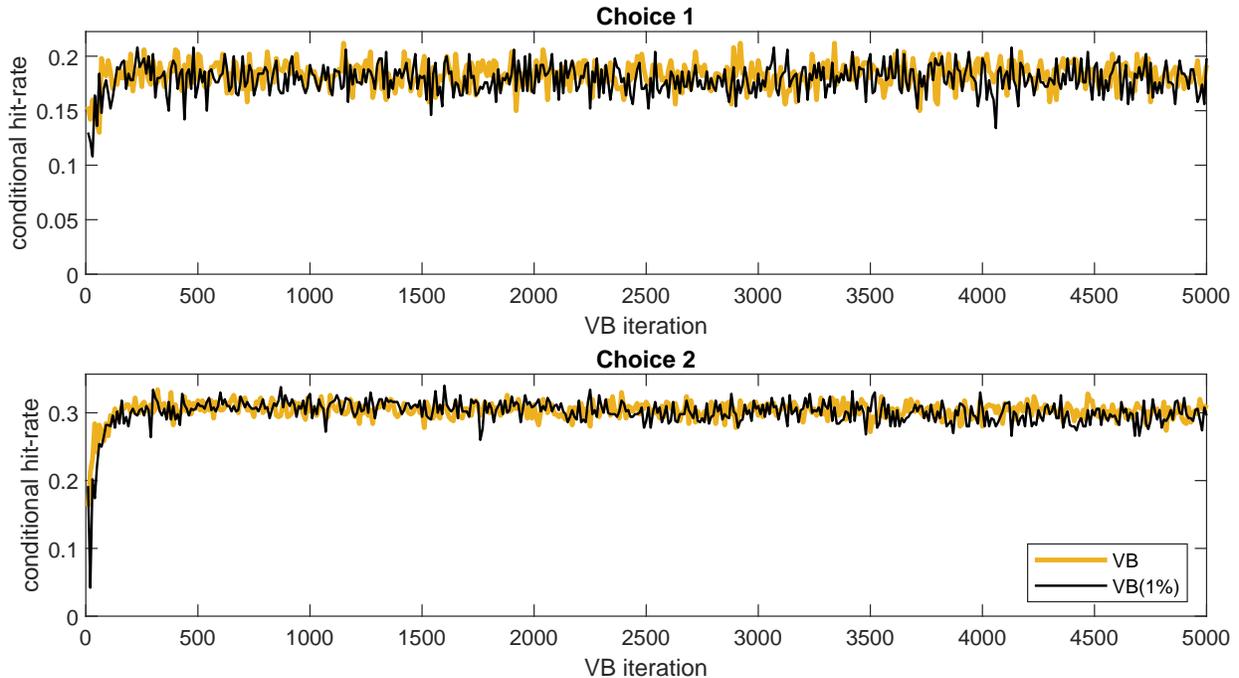}
 \fnote{This figure shows the conditional hit-rate for choice 1 and 2 in each tenth VB iteration, for VB and VB(1\%) by a yellow and black line, respectively. 
 }
\label{fig:conv}
\end{figure}

Figure~\ref{fig:conv} shows the conditional hit-rate for choice 1 and 2, in VB and VB(1\%) by a yellow and black line, respectively. 
The figure indicates that 5000 iterations are sufficient for convergence. The conditional hit-rates remain wiggly because they are constructed using an estimate for $\hat{p}_{\lambda^{[j]}}(\bm{Y}_i|X_i)$ that is based on a  $\bm\lambda^{[j]}$ that is updated using an estimate for the gradient. Therefore, the final estimate for the variational parameter $\bm\lambda$ is constructed as the average over the $\bm{\lambda}^{[j]}$ in the final 100 iterations. We find that VB converges in less iterations than VB(1\%). However, VB(1\%) has a faster convergence, as an average iteration takes 0.018 seconds, compared to 0.562 seconds per iteration in VB.

The computation time of VB(1\%), VB(10\%), and VB is 0.02, 0.11, and 0.77 hours respectively. Since MCMC takes 6.1 hours, this means that VB uses less than 14\% of the time required for MCMC, and the time can be further decreased with subsampling. 

\subsubsection{Parameter estimates}
\begin{figure}[tb!]
\caption{Posterior means and standard deviations in numerical experiment}
\centering
\includegraphics*[width=\textwidth,trim = 0 0 0 0]{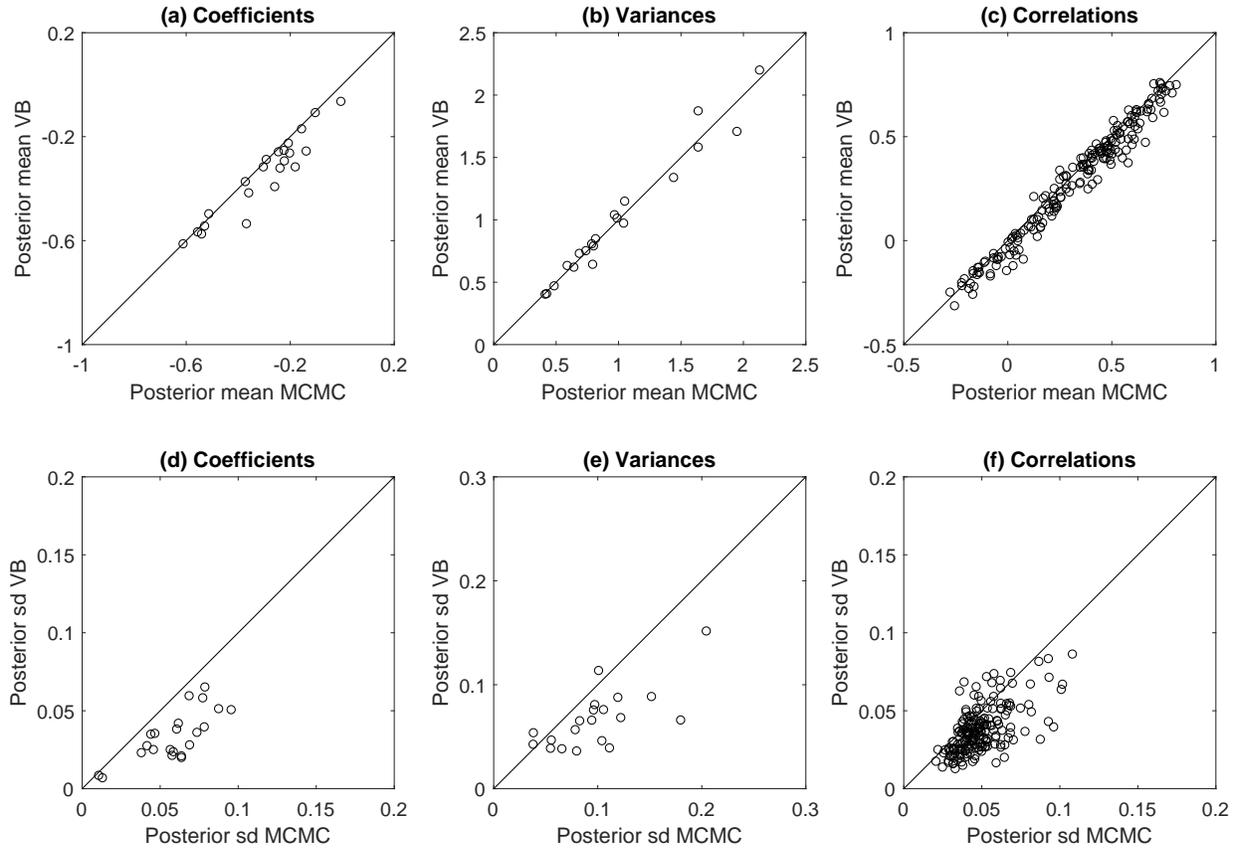}
 \fnote{Panels (a) to (c) present the estimated posterior means from MCMC (x-axis) against those from VB (y-axis). Panels (d) to (f) show corresponding plots for the posterior standard deviations. Panels (a) and (d) correspond to $\bm\beta$, Panels (b) and (e) correspond to the diagonal elements of $\Sigma$, and Panels (c) and (f) to the implied correlations.
 }
\label{fig:param}
\end{figure}

Second, we assess the accuracy of the posterior distribution for the parameters. 
Panels (a) to (c) in Figure~\ref{fig:param} show the VB against the MCMC posterior means. The closer the circles lie to the 45 degree line, the closer the VB posterior means are to those of MCMC. The VB estimates are scattered around the 45 degree line, which indicates that they are close to the exact posterior means. Panels (d) to (f) in Figure~\ref{fig:param} show that VB tends to underestimate the posterior standard deviation, which is a well-documented property of variational approximations \citep{blei2017variational,yu2021assessment}.

Online appendix~E compares the posterior means and standard deviations of VB(10\%) and VB(1\%) to those of MCMC. The posterior means of VB with subsampling are still scattered around the 45 degree line, but the deviations from this line slightly increase with smaller subsamples. This suggests that the reduction in computational costs induced by subsampling comes at the cost of a small loss in accuracy. VB with subsampling does not seem to underestimate the posterior standard deviation.

\begin{figure}[tb!]
\caption{Choice probabilities in the numerical experiment }
\centering
\includegraphics*[width=\textwidth,trim = 0 0 0 0]{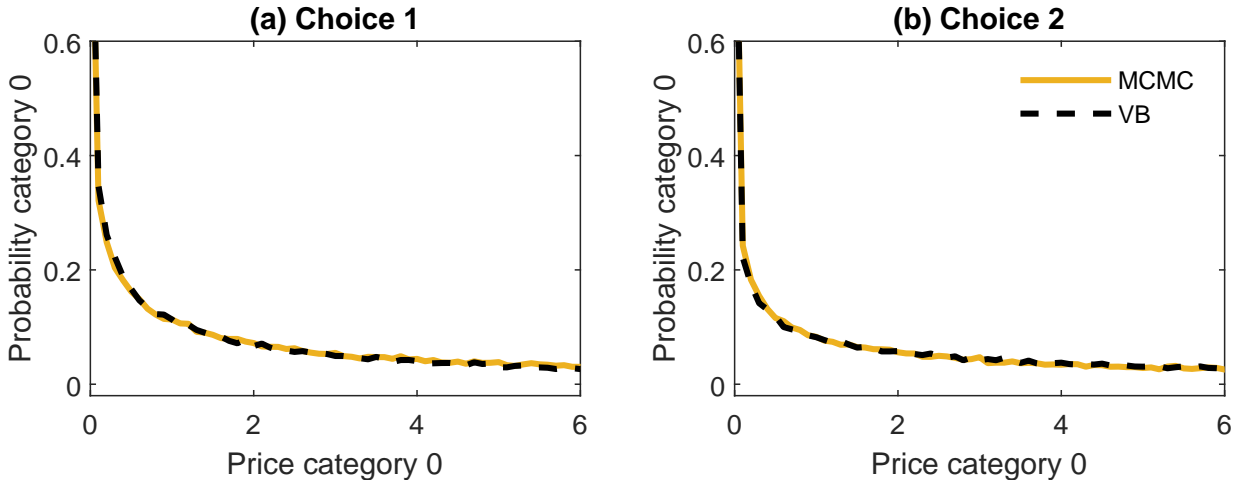}\\[3mm]
 \fnote{This figure shows the posterior choice probabilities of category 0 in choice 1 (Panel (a)) and choice 2 (Panel (b)) as a function of their price, with the prices of the other brands fixed at their mean. The solid yellow lines show the probabilities estimated with MCMC and the dashed black lines with VB.
 }
\label{fig:simul_prob}
\end{figure}

Although we have demonstrated the accuracy of VB at estimating the posterior of the parameters of the model, these parameter estimates themselves are hard to interpret and as such are not the key output from the model. Instead, the posterior choice probabilities are the quantity of interest in most empirical applications.
Figure~\ref{fig:simul_prob} shows the choice probability of one of the categories for choice 1 (Panel (a)) and for choice 2 (Panel (b)) as a function of their price, with the prices of the other categories fixed at their mean. The solid yellow lines correspond to the posterior probabilities of MCMC and the dashed black lines to VB. These lines are almost identical, and we find the same result for the other categories, and  when comparing VB(1\%) to MCMC. 
Hence we conclude that VB and VB with subsampling accurately estimate the posterior choice probabilities.

\subsubsection{Predictive accuracy}
\begin{figure}[tb!]
\caption{Log-scores and estimation time in numerical experiment with 10,000 observations}
\centering
\includegraphics*[width=\textwidth,trim = 0 0 0 0]{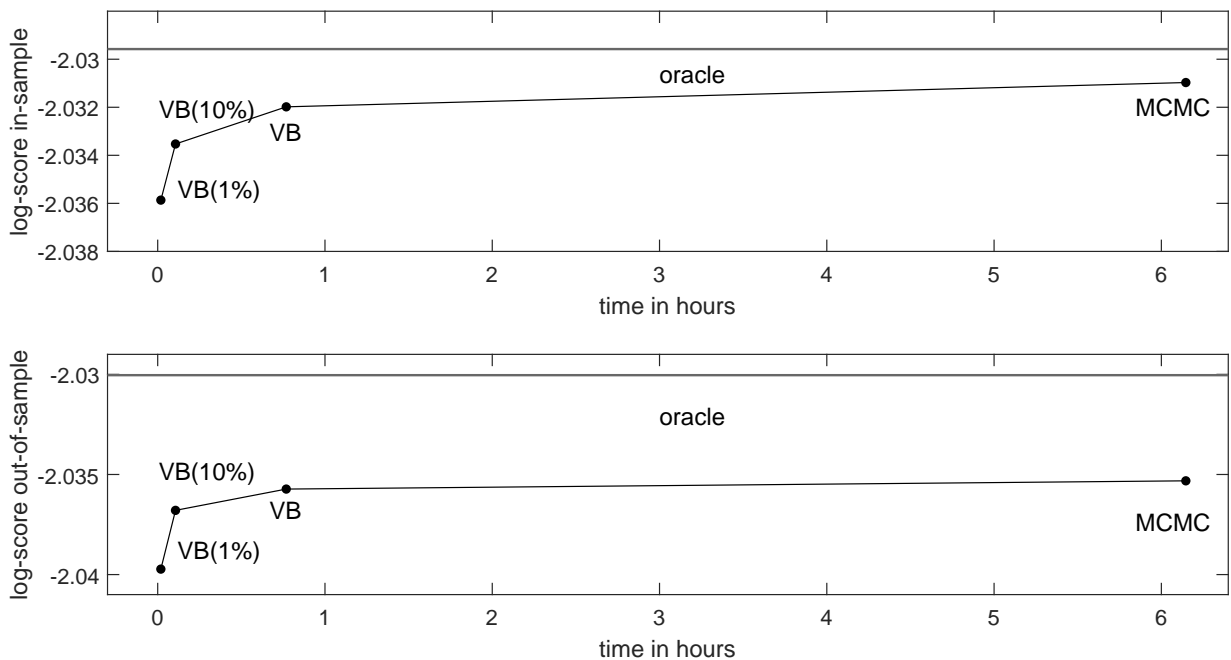}
 \fnote{This figure shows the log-score as defined in \eqref{eq:logscore} averaged over choices 1 and 2 against the estimation time, for VB with subsampling, VB, and MCMC. The first panel shows the in-sample log-score, and the second panel the out-of-sample log-score. The log-score of the oracle is computed using the true parameter values.}
\label{fig:ls_small}
\end{figure}

Third, we examine the predictive accuracy of our VB approach. The log-score can be used to assess the impact of subsampling in VB on the estimated model fit. Figure~\ref{fig:ls_small} shows the estimation time and in- and out-of-sample log-score across choices corresponding to VB with different subsampling sizes and MCMC. The log-score increases in the size of the subsample, with a big increase corresponding to subsampling with 1\% to 10\%. After 10\%, an increase in the subsample results in small gains in the log-scores.

Figure~\ref{fig:ls_small} shows that the impact of subsampling in VB on the estimated model fit is small compared to the gains in computational efficiency. For instance, VB(10\%) has an in- and out-of-sample log-score close to VB and MCMC. The differences between these three methods are small relative to the difference between MCMC and the log-score of the oracle: the log-score computed using the true parameter values. This is a striking result, as VB(10\%) is estimated in 7 minutes, VB takes 47 minutes, and MCMC more than 6 hours.

Table~\ref{tab:pred_small} provides a more detailed overview on the estimated model fit. The upper panel shows the in- and out-of-sample log-score and hit-rate for the first choice, and the lower panel for the second choice. These measures are higher in the second choice, indicating a stronger signal. MCMC performs better than VB on the log-scores, but does not always outperform the hit-rates of the different VB approaches. The average log-score of VB across choices declines with smaller subsamples. 
\begin{table}[tb!]
  \centering \small
  \caption{Log-score and hit-rate in numerical experiment with 10,000 observations}
  \begin{threeparttable}
    \begin{tabular}{llrrrrrrr}
    \toprule \toprule
    \multicolumn{9}{c}{Choice 1} \\
    \midrule
    Sample & Metric & \multicolumn{1}{l}{VB(1\%)} & \multicolumn{1}{l}{VB(10\%)} & \multicolumn{1}{l}{VB} & \multicolumn{1}{l}{VB-I} & \multicolumn{1}{l}{MCMC} & \multicolumn{1}{l}{Naive} & \multicolumn{1}{l}{Oracle} \\
    \midrule
in    & log-score &    -2.110 & -2.108 & -2.107 & -2.120 & -2.106 & -2.227 & -2.106 \\ 
    in    & hit-rate & 0.221  & 0.224  & 0.220  & 0.218  & 0.222  & 0.181  & 0.223  \\
    out   & log-score & -2.114 & -2.110 & -2.111 & -2.122 & -2.110 & -2.229 & -2.106 \\
    out   & hit-rate  & 0.232  & 0.231  & 0.230  & 0.230  & 0.233  & 0.186  & 0.234 \\ \midrule \midrule
    \multicolumn{9}{c}{Choice 2} \\
    \midrule
    Sample & Metric & \multicolumn{1}{l}{VB(1\%)} & \multicolumn{1}{l}{VB(10\%)} & \multicolumn{1}{l}{VB} & \multicolumn{1}{l}{VB-I} & \multicolumn{1}{l}{MCMC} & \multicolumn{1}{l}{Naive} & \multicolumn{1}{l}{Oracle} \\
    \midrule
    in    & log-score &-1.962 & -1.959 & -1.957 & -1.987 & -1.956 & -2.326 & -1.953 \\
    in    & hit-rate & 0.293  & 0.293  & 0.293  & 0.292  & 0.293  & 0.161  & 0.294  \\
    out   & log-score &-1.970 & -1.963 & -1.961 & -1.988 & -1.961 & -2.325 & -1.954 \\
    out   & hit-rate & 0.294  & 0.292  & 0.295  & 0.290  & 0.295  & 0.160  & 0.294 \\
    \bottomrule \bottomrule
    \end{tabular}%
\begin{tablenotes}
\footnotesize
\item This table shows the in- and out-of-sample log-scores and hit-rates, defined in respectively \eqref{eq:hitrate} and \eqref{eq:logscore}. Predictive densities are estimated with VB with subsampling, VB, VB with an identity covariance matrix, MCMC, a naive method in which the forecast equals the most frequently observed category, and the oracle that uses the true parameter values.
\end{tablenotes}
\end{threeparttable}
  \label{tab:pred_small}
\end{table}

VB in the MVMNP model can also be compared to VB in a choice model with an identity covariance matrix, VB-I. We find that VB-I is outperformed on all measures by VB. 
In general, all models perform better than the naive method on all metrics: the naive method sets the forecast equal the most frequently observed category in the data. The oracle, that uses the true parameter values to construct a forecast, corresponds to the highest log-scores, but does not always attain the highest hit-rates.

To study the robustness of the predictive performance of the model to the choice of the number of factors $p$, we repeat the predictive exercise for an increasing number of factors $p = 0,\dots,10$. Because the true DGP is generated from a full covariance matrix, smaller values of $p$ indicate a higher level of model misspecification.
We find that the increase in predictive performance is modest beyond $p=2$ factors, while the estimation time grows linearly with $p$. Details are deferred to online appendix E. 

\subsection{Results with one million observations}

The second experiment illustrates that VB with subsampling makes the estimation of the MVMNP model computationally feasible on big data sets.
Figure~\ref{fig:ls_small} shows that MCMC takes more than six hours with 10,000 observations. In practice, choice data sets may have much larger samples as we illustrate in Section~\ref{sec:application}. With one million observations, MCMC takes around eighteen days. 
These computational costs make MCMC impractical in many choice applications. VB takes almost three days, which is still a substantial computational cost. On the other hand, VB(1\%) takes less than 50 minutes to be implemented.

An equally time efficient approach that could be implemented instead of subsampling VB, would be to consider VB on a random subsample of the data. This approach does not make use of the complete data set, and as such can be suboptimal in terms of predictive accuracy.
To show this, we apply VB to a subsample of 10,000 observations, which requires approximately the same computation time as VB(1\%) on a million observations.
Table~\ref{tab:pred_large} shows the log-scores and hit-rates of the two approaches on the same out-of-sample observations, and the average log-score of VB(1\%) is indeed higher than that of VB.


\begin{table}[tb!]
  \centering \small
  \caption{Out-of-sample log-score and hit-rate in large numerical experiments}
  \begin{threeparttable}
    \begin{tabular}{lrrrr}
    \toprule \toprule
    N     & \multicolumn{2}{c}{10,000} & \multicolumn{2}{c}{1,000,000} \\
    Method & \multicolumn{2}{c}{VB}  & \multicolumn{2}{c}{VB(1\%)} \\
    \cline{2-3}\cline{4-5}
    Metric & \multicolumn{1}{l}{log-score} & \multicolumn{1}{l}{hit-rate} &  \multicolumn{1}{l}{log-score} & \multicolumn{1}{l}{hit-rate} \\
    Choice 1 &-2.111 & 0.230 & -2.107 & 0.232 \\
    Choice 2 &-1.961 & 0.295 & -1.960 & 0.293 \\
    Average & -2.036 & 0.263 & -2.033 & 0.263\\
    \bottomrule \bottomrule
    \end{tabular}%
\begin{tablenotes}
\footnotesize
\item This table shows the out-of-sample log-scores and hit-rates, defined in respectively \eqref{eq:hitrate} and \eqref{eq:logscore}. The parameters are estimated with VB on $N=10,000$ observations from the data generating process and VB with 1\% subsampling on $N=1,000,000$. The predictive densities are estimated on the same out-of-sample of 10,000 observations for all three methods.
\end{tablenotes}
\end{threeparttable}
  \label{tab:pred_large}
\end{table}

\section{Empirical application}\label{sec:application}
To illustrate our VB method with real data, we fit a multinomial probit model to two consumer choice data sets with different dimensions. First, Section~\ref{sec: small} employs a commonly used traditional data set on laundry detergent brand purchases with a few thousand observations. Second, Section~\ref{sec: big} uses a modern data set on pasta brand purchases with more than one million observations. We discuss the posterior choice probabilities and the predictive performance of VB and MCMC, and defer the results on the posterior parameter distributions to online appendix~F. The implementation and prior settings of the proposed methods are discussed in Section~\ref{sec:design}. We randomly allocate 80\% of the observations for estimation of the model, and the remaining 20\% are employed for out-of-sample evaluation.

\subsection{Small data set with laundry detergent purchases}\label{sec: small}
This section uses a small choice data set to illustrate on real data that VB is several times faster than MCMC, yet produces similar choice probabilities and predictive accuracy. The data contains 2657 purchases of six brands of laundry detergents and the log price per ounce of each brand. The data set is described in detail by \citet{chintagunta1998empirical} and available in \citet{imai2005mnp}. We follow \citet{imai2005bayesian}, \citet{burgette2019symmetric}, and \citet{loaiza2021scalable} by fitting multinomial probit models with an intercept and the log price for each brand.

\begin{figure}[tb!]
\caption{Purchase probabilities for two detergent brands}
\centering
\includegraphics*[width=\textwidth,trim = 0 0 0 0]{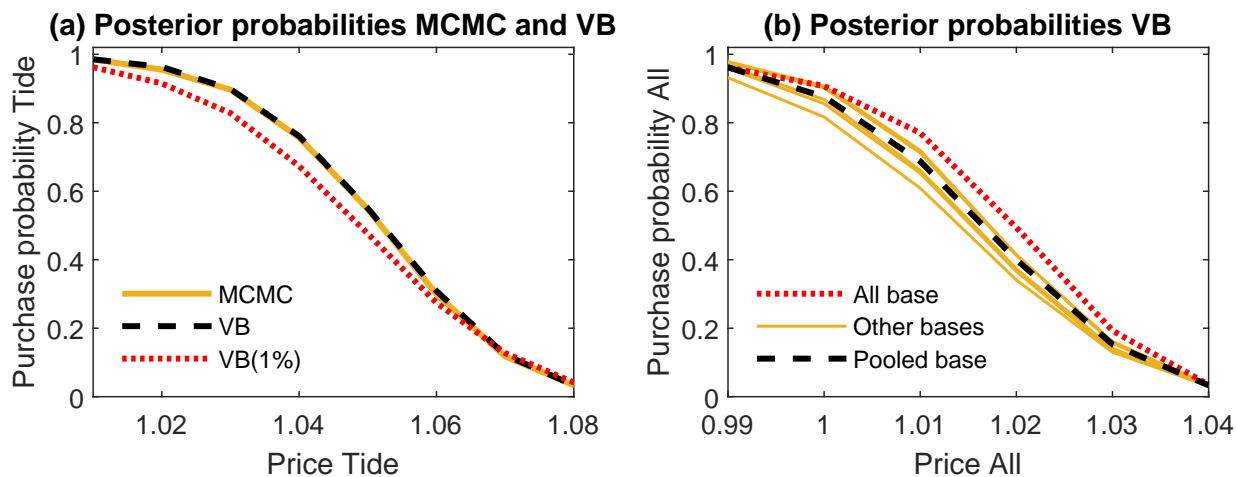}\\[3mm]
\fnote{This figure shows the posterior purchase probabilities of a detergent brand as a function of its price, with the prices of the other brands fixed at their mean. Panel (a) shows the probability of buying the most popular brand `Tide' estimated with MCMC (solid yellow line), VB (dashed black line), and VB(1\%) (dotted red line). Panel (b) shows the probability of buying the least popular brand `All' estimated with VB and `All' as base category (dotted red line), the other five base categories (solid yellow black lines), and averaged across the six probabilities corresponding to each base category (dashed black line). }
\label{fig:detergent_prob}
\end{figure}

We find that the posterior purchase probabilities of VB and MCMC are similar. Panel~(a) in Figure~\ref{fig:detergent_prob} shows the probability of buying the most popular brand as a function of its price, with the prices of the other brands fixed at their mean. The solid yellow line corresponds to the posterior probabilities of MCMC, the dashed black line to VB, and the dotted red line to VB(1\%). VB produces posterior probabilities that are almost identical to MCMC, and VB(1\%) only shows small differences compared to MCMC. We also find negligible differences between VB and MCMC, and small differences between VB(1\%) and MCMC, for the purchase probabilities for the other five brands.

VB also attains similar predictive accuracy to MCMC. Table~\ref{tab:forecast_small} shows that the in- and out-of-sample log-scores of VB and MCMC are almost identical. The log-scores slowly decrease in the order of subsampling in VB. The hit rates do not seem to be very sensitive to the approximations by VB, or in the VB with subsampling. Both the log-scores and hit-rates of VB(10\%) are relatively close to MCMC, especially when we consider the difference in these metrics between MCMC and the naive forecasting method, in which the forecast equals the most frequently observed category in the data. The fact that VB with an identity covariance matrix results in lower log-scores than VB with a full covariance matrix, indicates that the correlations across the latent utilities matter in this application. 

\begin{table}[tb!]
  \centering
  \caption{Log-score and hit-rate for laundry detergent application}
  \begin{threeparttable}
    \begin{tabular}{llrrrrrr}
    \toprule \toprule
          Sample & Metric       & \multicolumn{1}{l}{VB(1\%)} & \multicolumn{1}{l}{VB(10\%)} & \multicolumn{1}{l}{VB} & \multicolumn{1}{l}{VB-I} & \multicolumn{1}{l}{MCMC} & \multicolumn{1}{l}{Naive} \\
          \midrule
    in    & log-score & -1.345 & -1.333 & -1.332 & -1.352 & -1.332 & -1.639 \\
    in    & hit-rate & 0.490 & 0.500 & 0.501 & 0.505 & 0.494 & 0.270 \\
    out   & log-score & -1.388 & -1.388 & -1.385 & -1.407 & -1.383 & -1.657 \\
    out   & hit-rate & 0.507 & 0.495 & 0.503 & 0.501 & 0.499 & 0.239 \\
    \midrule
          & time (seconds) & 15.371 & 37.529 & 80.719 & 77.262 & 720.304 &  \\
          \bottomrule \bottomrule
    \end{tabular}%
\begin{tablenotes}
\footnotesize
\item This table shows the in- and out-of-sample log-scores and hit-rates, defined in respectively \eqref{eq:logscore} and \eqref{eq:hitrate}. The final row shows the estimation time in seconds. Predictive densities are estimated with VB with subsampling, VB, VB with an identity covariance matrix, MCMC, and a naive method in which the forecast equals the most frequently observed category.
\end{tablenotes}
\end{threeparttable}
  \label{tab:forecast_small}
\end{table}

Although the differences in parameter estimates and predictive accuracy are minimal, VB is more than eight times faster than MCMC. The final row of Table~\ref{tab:forecast_small} shows that MCMC takes 720 seconds while VB only takes 81 seconds. This computation time can be further reduced by subsampling. VB(10\%) takes 38 seconds, as it required 10,000 rather than 5,000 SGA iterations to converge, and VB(1\%) only 16 seconds.

Due to its low computational costs, VB is well suited to estimate multiple prior specifications to study the robustness of the posterior results to the choice of base category. For instance, \citet{burgette2021symmetric} show that the posterior choice probabilities of Bayesian MNP models can depend on the base category specification. 
Panel~(b) in Figure~\ref{fig:detergent_prob} shows the probability of buying the least popular detergent brand `All' estimated with VB. The dotted red line uses brand `All' as the base category, and the solid yellow lines correspond to the five other base category specifications. The lines are different, and `All' as base category results in substantially higher purchase probabilities than with other base categories.      
The differences are less pronounced for choice probabilities of more popular brands. Setting the number of factors equal to the number of brands shows the same base category sensitivities, which is in line with the findings in \citet{burgette2021symmetric} who specify a full covariance matrix. 
As a robust alternative for estimating the choice probabilities, the posterior probabilities corresponding to different base category specifications can be pooled. The dashed black line in Panel~(b) in Figure~\ref{fig:detergent_prob} shows the average purchase probability across all specifications.

\subsection{Large data set with pasta purchases}\label{sec: big}
This section shows that our approach can be scaled to real data with many observations. 
We use more than one million purchases from ten pasta brands in a consumer choice data set made available by Dunnhumby\footnote{https://www.dunnhumby.com/source-files/} as ``Carbo-Loading: A Relational Database". From this data set, we select the purchases of pasta brands, excluding the private labels, that do not involve coupons. Since the brands with a small purchase volume are of less interest to a marketing manager, we focus on the 96.782\% of purchases that corresponds to the ten top-selling pasta brands. The final sample contains 1,070,436 observations and the purchase frequencies vary from 6,280 to 316,018. 

We consider the same MNP models as with the small data set on laundry detergent purchases, also including an intercept and the log price for each brand. 
We define the log price per ounce of each brand in the same way as, for instance, \citet{allenby1991quality} and \citet{loaiza2021scalable}. The Dunnhumby data set only contains the amount of dollar spent on a product at purchase dates. We impute the prices for brands that are not sold on a certain purchase date by taking the mean of the observed prices of a specific brand on the nearest date in the same week. In cases where there is no purchase record in the same week, we take the most recent observed price. 

\begin{figure}[tb!]
\caption{Purchase probabilities for two pasta brands}
\centering
\includegraphics*[width=\textwidth,trim = 0 0 0 0]{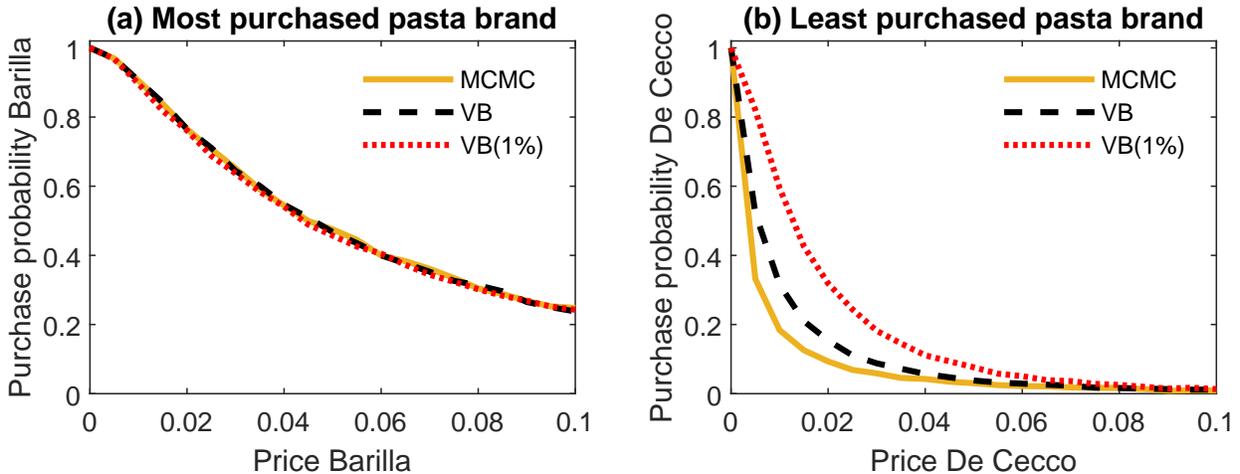}\\[3mm]
 \fnote{This figure shows the posterior purchase probabilities of the most (Barilla, Panel (a)) and least (De Cecco, Panel (b)) popular pasta brands as a function of their price, with the prices of the other brands fixed at their mean. The solid yellow line shows the probabilities estimated with MCMC, the dashed black line with VB, and the dotted red line with VB(1\%). 
 }
\label{fig:pasta_prob}
\end{figure}

Figure~\ref{fig:pasta_prob} shows the purchase probabilities as a function of its price for the most popular pasta brand in Panel (a) and the least popular pasta brand in Panel (b). The solid yellow line corresponds to the posterior probabilities of MCMC, the dashed black line to VB, and the dotted red line to VB(1\%). These lines are almost identical for the most purchased pasta brand. For the least purchased pasta brand, the probabilities of MCMC are different from the probabilities of VB(1\%) and VB, although the difference with the latter is small. In general, VB produces posterior probabilities that are accurate for the relatively large pasta brands, and lose some accuracy for brands with a small number of purchases.

Table~\ref{tab:forecast_big} reports the predictive performance measures. Based on the in-sample and out-of-sample log-scores and hit-rates, VB and VB(10\%) show almost no loss in accuracy relative to MCMC. The table also indicates that the difference in predictive performance between MCMC and VB(1\%) is small  compared to the difference between MCMC and the naive method.
We also find that, irrespective of the subsampling size, VB outperforms VB-I, which emphasises the importance of taking correlations into account.

Table~\ref{tab:forecast_big} also shows that VB makes the estimation of multinomial probit models feasible on real choice data sets with many observations. 
The final row shows that implementation of VB takes more than fifteen hours, which is a  fraction of the 92 hours of MCMC. VB can even further reduce the computation time with subsampling, with VB(10\%) only taking around 1.5 hours, and VB(1\%) around eleven minutes. 

\begin{table}[tbh]
  \centering
  \caption{Log-score and hit-rate for pasta application}
  \begin{threeparttable}
    \begin{tabular}{llrrrrrr}
    \toprule \toprule
          Sample & Metric       & \multicolumn{1}{l}{VB(1\%)} & \multicolumn{1}{l}{VB(10\%)} & \multicolumn{1}{l}{VB} & \multicolumn{1}{l}{VB-I} & MCMC & \multicolumn{1}{l}{Naive} \\
          \midrule
    in    & log-score & -1.777 & -1.773 & -1.773 & -1.786 &-1.773 &-1.845 \\
    in    & hit-rate & 0.397 & 0.398 & 0.398 & 0.379 &0.399 &0.295 \\
    out   & log-score & -1.777 & -1.774 & -1.774 & -1.786 &-1.774 &-1.842 \\
    out   & hit-rate & 0.396 & 0.397 & 0.397 & 0.377 & 0.397 &0.296 \\
    \midrule
          & time (hours) & 0.175 & 1.498 & 15.253 & 11.009 &92.291&  \\
          \bottomrule \bottomrule
    \end{tabular}%
\begin{tablenotes}
\footnotesize
\item This table shows the in- and out-of-sample log-scores and hit-rates, defined in respectively \eqref{eq:logscore} and \eqref{eq:hitrate}. The final row shows the estimation time in seconds. Predictive densities are estimated with VB with subsampling, VB, VB with an identity covariance matrix, MCMC and a naive method in which the forecast equals the most frequently observed category in the data.
\end{tablenotes}
\end{threeparttable}
  \label{tab:forecast_big}
\end{table}

\section{Conclusion}\label{sec: conclusion}
Multinomial probit models are widely used for analyzing choice behavior. The main benefit of the model is the specification of the covariance matrix of the latent utilities. 
To accurately estimate the covariance parameters from a single categorical dependent variable, a large number of observations is required.  
Choice data sets with many observations are nowadays widely available. For instance, scanner data as used in the empirical application in this paper have records of millions of transactions. 
%
However, MCMC methods that are currently used for parameter estimation are computationally costly. 

This paper proposes a variational Bayes method that employs the conditional posterior of the latent utilities as a part of the variational family. This allows for accurate approximations to the exact posterior. The method is faster than MCMC with moderately sized data sets, and is scalable to large-scale data in which MCMC estimation is infeasible. 

Numerical experiments and an empirical application to a laundry detergent choice set demonstrate that our approach produces accurate approximating densities to the MCMC exact posterior densities, while only requiring a small fraction of the MCMC computation time. The computational cost for our approach can be further reduced by considering subsampling methods inside the stochastic gradient ascent algorithm, with small impact in its predictive accuracy relative to MCMC.

The new method improves the applicability of the multinomial probit model to modern choice data sets. We illustrate the potential of the new approach in large samples by applying it to a pasta choice data set that consists of more than one million observations.



\bibliographystyle{apalike}

\bibliography{mvmnp}
\clearpage
\bigskip
\newpage
\noindent
\setcounter{page}{1}
\begin{center}
	{\bf \Large{Online appendix for ``Fast variational Bayesian methods for
multinomial probit models''}}
\end{center}

\vspace{10pt}

\setcounter{figure}{0}
\setcounter{table}{0}
\setcounter{section}{0}
\renewcommand{\thetable}{A\arabic{table}}
\renewcommand{\thefigure}{A\arabic{figure}}

\noindent
This online appendix has six parts:

\begin{itemize}
	\item[] {\bf Part~A}: Specification of the prior distributions.
	\item[] {\bf Part~B}: Description of the MCMC sampling algorithm.
	\item[] {\bf Part~C}: Details of the implementation of VB in the MVMNP model.
	\item[] {\bf Part~D}: Details of the implementation of VB in the MVMNP model with identity covariance matrix.
	\item[] {\bf Part~E}: Additional results numerical experiments.
	\item[] {\bf Part~F}: Additional results empirical applications.
\end{itemize}
\newpage

\appendix

\section{Prior specification}\label{A:prior}
The prior distribution is defined as $p(\bm{\theta}) = p(\bm\beta)\prod_{k=1}^{K}p(\bm\kappa_k)$, with $\bm{\beta}\sim N(\bm{0}_r, \frac{1}{10}I_r)$, $p(\bm\kappa_k)=\prod_{l=1}^{n_k-1}p(\kappa_{kl})$, and $p({\kappa}_{kl}) = \phi_1\left\{t_{\hat{\eta}_{kl}}\left[\frac{G\left(\kappa_{kl}\right)-\hat{\mu}_{kl}}{\hat{\tau}_{kl}}\right]\right\}t_{\hat{\eta}_{kl}}'\left[\frac{G\left(\kappa_{kl}\right)-\hat{\mu}_{kl}}{\hat{\tau}_{kl}}\right]\frac{1}{\hat{\tau}_{kl}}G'\left(\kappa_{kl}\right)$. The function $t_{\hat{\eta}_{kl}}$ denotes the Yeo-Johnson transformation, and the hyperparameters $\hat{\mu}_{kl}$, $\hat{\tau}_{kl}$, and $\hat{\eta}_{kl}$ are calibrated for each choice $k$ at a time using Algorithm 1 in \cite{loaiza2021scalable}.
%
    %

The algorithm calibrates the prior for $\bm\kappa_k$ using the following prior on $\bm\psi$:
\begin{align}
    \bm\psi_k &= \frac{\sqrt{J_k}}{\lVert\ddot{\bm\psi_k}\lVert}\ddot{\bm\psi}_k,\label{eq: psiprior0}\\
    p(\ddot{\bm\psi}_k|\bm\theta) &= \prod_{j=1}^{J}\left[p({\ddot{d}}_{kj}|\nu,s)\prod_{l=1}^{J}p({\ddot{B}}_{kjl}|\sigma^2_{B})\right],\label{eq: psiprior1}\\
    {\ddot{B}_{kjl}|\sigma_{B}^2}&\sim N({\mu_{B}},\sigma_{B}^2), \text{ if } j\neq l,\label{eq: psiprior3}\\
    {\ddot{B}_{kjj}|\sigma_{B}^2}&\sim N({\mu_{B}},\sigma_{B}^2)I(\ddot{B}_{kjj}>0),\\
    {\ddot{d}}_{kj}^2|\nu,s&\sim\text{Inverse-Gamma}\left(\nu,s\right),\label{eq: psiprior4}
\end{align}
where $\nu$ and $s$ denote the shape and rate parameters of the Inverse-Gamma distribution. We calibrate $\mu_{B}$ to obtain an implied prior mean for $\Sigma$ that equals the equicorrelated covariance matrix $\frac{1}{2}(I_J + \iota_J \iota_J^\top)$ as in \cite{loaiza2021scalable}, and set $\sigma_{B}^2=1$, $\nu=5$, and $s=\nu-1$.

The angles ${\kappa}_{kl}$ are transformed to the real line as 
\begin{align}\label{eq:xi}
    \xi_{kl} = \begin{cases} G\left(\kappa_{kl}\right)=\Phi^{-1}\left(\frac{\kappa_{kl}}{\pi}\right) & \text{ if } l<n_k-J_k+1,\\
    G\left(\kappa_{kl}\right)=\Phi^{-1}\left(\frac{\kappa_{kl}}{\pi/2}\right) & \text{ if } n_k-J_k+1\le l \le n_k-1.\end{cases}
\end{align}
After transforming, we use the Jacobian of the transformation to obtain the prior density
\begin{align}
    p({\xi}_{kl}) =  \phi_1\left[t_{\hat{\eta}_{kl}}\left(\frac{{\xi}_{kl}-\hat{\mu}_{kl}}{\hat{\tau}_{kl}}\right)\right]t_{\hat{\eta}_{kl}}'\left(\frac{{\xi}_{kl}-\hat{\mu}_{kl}}{\hat{\tau}_{kl}}\right)\frac{1}{\hat{\tau}_{kl}},
\end{align}
and the full prior density for $\bm{\theta} = \left(\bm{\beta}^\top,\bm{\xi}^\top\right)^\top$ can be written as
\begin{equation}
    p(\bm{\theta}) = p(\bm{\beta})\prod_{k=1}^K\prod_{l=1}^{n_k-1}p({\xi}_{kl}).
\end{equation}


\section{Monte Carlo Markov Chain sampling scheme}\label{A:mcmc}
Define $L_k=\sum_{l=1}^k J_l$. 
Define the $L_K$-dimensional vectors $Y_i=(Y_{i1}^\top,\dots,Y_{iK}^\top)^\top$ and  $Z_i=(Z_{i1}^\top,\dots,Z_{iK}^\top)^\top$, and the $L_K \times q$ matrix $X_i=(X_{i1}^\top,\dots,X_{iK}^\top)^\top$. Define $Y=(Y_{1}^\top,\dots,Y_{N}^\top)^\top$, $Z=(Z_{1}^\top,\dots,Z_{N}^\top)^\top$, and $X=(X_{1}^\top,\dots,X_{N}^\top)^\top$. Samplings steps for $\beta$ and $Z$ are standard and also discussed in, for instance, \citet{zhang2008bayesian}. 

\subsection{Conditional posterior $\bm\beta$}
Generate from $\bm\beta|Z,\Sigma,X$: The coefficients $\bm\beta$ are generated from
\begin{align}
    \bm\beta|Z,\Sigma,X\sim\mathcal{N}(\bm{\bar{b}},\bar{B}^{-1}),
\end{align}
with $\bar{B}={X^*}^\top{X^*}+B$ and $\bm{\bar{b}}=\bar{B}^{-1}{X^*}^\top{Z^*}$, where $X^*=(X_1^\top C,\dots,X_N^\top C)^\top$ and $Z^*=(Z_1^\top C,\dots,Z_N^\top C)^\top$, with $\Sigma^{-1}=CC^\top$.

\subsection{Conditional posterior $Z$}\label{A:condz}
Generate from $Z|\beta,\Sigma,Y,X$. To generate from the latent utilities we employ the truncated normal distributions,
    \begin{align}
        Z_{ikj}&\sim {N}^+_{\max(Z_{ik}^{(j)},0)}(\bar{\mu}_{ikj},\bar{\Sigma}_{ikj}),  \text{ if }  Y_{ik}=j,\\
        Z_{ikj}&\sim{N}^-_{\max(Z_{ik}^{(j)},0)}(\bar{\mu}_{ikj},\bar{\Sigma}_{ikj}),  \text{ if }  Y_{ik}\neq j,
    \end{align}
where ${N}^+_{a}(\mu,\sigma^2)$ and ${N}^-_{a}(\mu,\sigma^2)$ represent a normal distribution with mean $\mu$ and variance $\sigma^2$ truncated from below or above by $a$, respectively, and  $Z_{ik}^{(j)}=(Z_{ik1},\dots,Z_{ikj-1},Z_{ikj+1},\dots,Z_{ikJ_k})^\top$. The conditional mean and variance of $Z_{ikj}$ given $Z_i^{(kj)}=(Z_{i1}^\top,\dots,Z_{ik-1}^\top,Z_{ik}^{(j)\top},Z_{ik+1}^\top,\dots,Z_{iK}^\top)^\top$, are defined as
\begin{align}
    \bar{\mu}_{ikj} &= X_{ikj}\beta+\Sigma_{kj(kj)}\Sigma_{(kj)(kj)}^{-1}(Z_i^{(kj)}-X_i^{(kj)}\beta),\\
    \bar{\Sigma}_{ikj} &= \Sigma_{kjkj}-\Sigma_{kj(kj)}\Sigma_{(kj)(kj)}^{-1}\Sigma_{(kj)kj},
\end{align}
where $X_i^{(kj)}$ is defined in the same way as $Z_i^{(kj)}$, $\Sigma_{kjkj}$ is the element in row and column number $L_{k-1}+j$ of $\Sigma$, $\Sigma_{kj(kj)}$ ($\Sigma_{(kj)kj}$) is row (column) $L_{k-1}+j$ of $\Sigma$ without column (row) $L_{k-1}+j$, and $\Sigma_{(kj)(kj)}$ equals $\Sigma$ after removing row and column number $L_{k-1}+j$. 

\subsection{Conditional posterior $\bm\kappa$}\label{A:condkappa}
Sampling of the parameters $\bm\kappa$ is obtained via blocked random walk Metropolis-Hastings steps. 
At the start of each iteration, randomly allocate the elements of $\bm\kappa$ into $G$ parameter blocks, $\bm\kappa_{b_1},\dots,\bm\kappa_{b_G}$, of five elements each. 
Note that in this appendix each index $i$ in $\bm\kappa_i$ refers to a block of random elements of $\bm\kappa$ and not to the choice-specific vector of angles as in Section~\ref{sec:identification}.
For $g = 1,\dots,G$, generate a draw $\bm\kappa_{b_g}^{\text{new}}$ from the proposal density,
\begin{align}
q(\bm\kappa_{b_g}|\bm\kappa_{b_g}^{\text{old}}) = \prod_{l=1}^5\frac{\phi_1\left(\kappa_{b_g,l};\kappa_{b_g,l}^{\text{old}},\sigma_{g_l}^2\right)}{\Phi_1\left(\text{up}_{g_l};\kappa_{b_g,l}^{\text{old}},\sigma_{g_l}^2\right)-\Phi_1\left(\text{low}_{g_l};\kappa_{b_g,l}^{\text{old}},\sigma_{g_l}^2\right)},
\end{align}
where $\kappa_{b_g,l}$ is the $l$th element of $\bm\kappa_{b_g}$. The constants $\text{low}_{g_l}$ and $\text{up}_{g_l}$ denote the lower and upper bounds of $\kappa_{b_g,l}$. The proposal parameters $\sigma_{g_l}^2$ are set adaptively to target acceptance rates between $15\%$ and $30\%$. We accept $\bm\kappa_{b_g}^{\text{new}}$  with probability
\begin{align}
	\alpha = \min\left(1,\frac{p(\bm\kappa_{b_g}^{\text{new}}|\bm z,X,\left\{\bm\theta\backslash\bm\kappa_{b_g}\right\})q(\bm\kappa_{b_g}^{\text{old}}|\bm\kappa_{b_g}^{\text{new}})}{p(\bm\kappa_{b_g}^{\text{old}}|\bm z,X,\left\{\bm\theta\backslash\bm\kappa_{b_g}\right\})q(\bm\kappa_{b_g}^{\text{new}}|\bm\kappa_{b_g}^{\text{old}})}\right),
\end{align}
where
\begin{align}
	p(\bm\kappa_{b_g}|\bm z,X,\left\{\bm\theta\backslash\bm\kappa_{b_g}\right\})\propto p(\bm\kappa_{b_g})p(\bm z|X,\bm{\theta}).
\end{align}
The expression $\left\{\bm\theta\backslash\bm\kappa_{b_g}\right\}$ denotes the subtraction of the subset $\bm\kappa_{b_g}$ from $\bm\theta$.

\section{VB in the MVMNP model}\label{A:VI}
VB in the MVMNP model requires an unbiased estimate of the gradient $${\nabla_{\theta}\log g}\left(\bm{\theta},\bm{z}\right) = \left(\nabla_{\beta}\log g(\bm{\theta},\bm{z})^\top,\nabla_{\xi}\log g(\bm{\theta},\bm{z})^\top\right)^\top,$$ where $\bm{\theta} = \left(\bm{\beta}^\top,\bm{\xi}^\top\right)^\top$, $\bm{\xi} = \left(\bm{\xi}_1^\top,\dots,\bm{\xi}_K^\top\right)^\top$, $\bm{\xi}_k = \left(\xi_{k1},\dots,\xi_{k,n_k-1}\right)^\top$, and  $\xi_{kl}$ is defined in \eqref{eq:xi}. The function $\log g(\bm{\theta},\bm{z})^\top$ can be written as 
\begin{equation}
 \log g(\bm{\theta},\bm{z}) = \log p(\bm{y}|\bm{z})+ \log p(\bm{\theta})+\sum_{i=1}^{N}\log \phi_{J}\left(\bm{z}_i;X_i\bm{\beta},\Sigma(\bm{\xi})\right),
\end{equation}
where 
\begin{equation}
 \log p(\bm{\theta}) = \log p(\bm{\beta})+\sum_{k=1}^{K}\sum_{l=1}^{n_k-1}\log p(\xi_{kl}),
\end{equation}
and $\Sigma(\bm{\xi})$ reflects the fact that the covariance matrix now depends on $\bm{\xi}$.

The estimate can be constructed on a set $A\subset\{1,\dots,N\}$ of $M$ indexes sampled at random and without replacement:
\begin{align}
    {\nabla_{\beta}\log g(\bm{\theta},\bm{z}_A)}& = \nabla_{\beta}\log p(\bm{\theta})+\frac{N}{M}\sum_{i\in A}\nabla_{\beta}\log \phi_{J}\left(\bm{z}_i;X_i\bm{\beta},\Sigma(\bm{\xi})\right),\\
    {\nabla_{\xi}\log g(\bm{\theta},\bm{z}_A)}& = \nabla_{\xi}\log p(\bm{\theta})+\frac{N}{M}\sum_{i\in A}\nabla_{\xi}\log \phi_{J}\left(\bm{z}_i;X_i\bm{\beta},\Sigma(\bm{\xi})\right),
\end{align}
which boils down to the exact evaluation of ${\nabla_{\theta}\log g}\left(\bm{\theta},\bm{z}\right)$ when $N=M$ and there is no subsampling. 
We show how to evaluate the terms $\nabla_{\beta}\log p(\bm{\theta})$, $\nabla_{\xi}\log p(\bm{\theta})$, $\nabla_{\beta}\log \phi_{J}\left(\bm{z}_i;X_i\bm{\beta},\Sigma(\bm{\xi})\right)$ and $\nabla_{\xi}\log \phi_{J}\left(\bm{z}_i;X_i\bm{\beta},\Sigma(\bm{\xi})\right)$ below. Throughout we use $\bm{\eta}_i = \bm{z}_i-X_i\bm{\beta}$.

\subsection{Computing $\nabla_{\beta}\log p(\bm{\theta})$}\label{A:dbetaprior}
\begin{align}
\nabla_{\beta}\log p(\bm{\theta}) = -10\bm{\beta}.
\end{align}

\subsection{Computing $\nabla_{\xi}\log p(\bm{\theta})$}
To compute this gradient we know that
$\nabla_{\xi}\log p(\bm{\theta}) = \left(\nabla_{\xi_1}\log p(\bm{\theta})^\top,\dots,\nabla_{\xi_K}\log p(\bm{\theta})^\top\right)^\top$, where 
$$\nabla_{\xi_k}\log p(\bm{\theta}) = \left(\frac{\partial \log p(\xi_{k,1})}{\partial \xi_{k,1}},\dots,\frac{\partial \log p(\xi_{k,n_k-1})}{\partial \xi_{k,n_k-1}}\right)^\top,$$
and 
$$\frac{\partial \log p(\xi_{kl})}{\partial \xi_{kl}} = -\frac{1}{\hat{\tau}_{kl}}t_{\hat{\eta}_{kl}}\left[\frac{\xi_{kl}-\hat{\mu}_{kl}}{\hat{\tau}_{kl}}\right]t_{\hat{\eta}_{kl}}'\left[\frac{\xi_{kl}-\hat{\mu}_{kl}}{\hat{\tau}_{kl}}\right]+\frac{1}{\hat{\tau}_{kl}}\frac{t_{\hat{\eta}_{kl}}''\left[\frac{\xi_{kl}-\hat{\mu}_{kl}}{\hat{\tau}_{kl}}\right]}{t_{\hat{\eta}_{kl}}'\left[\frac{\xi_{kl}-\hat{\mu}_{kl}}{\hat{\tau}_{kl}}\right]}.$$
Closed-form expressions for $t_{\hat{\eta}_{kl}}(\cdot)$, $t_{\hat{\eta}_{kl}}'(\cdot)$ and $t_{\hat{\eta}_{kl}}''(\cdot)$ are provided in Table 1 in \cite{smith2020high} for the Yeo-Johnson transformation.

\subsection{Computing $\nabla_{\beta}\log \phi_{J}\left(\bm{z}_i;X_i\bm{\beta},\Sigma(\bm{\xi})\right)$}\label{A:dbetaZ}
\begin{align}
    \nabla_{\beta}\log \phi_{J}\left(\bm{z}_i;X_i\bm{\beta},\Sigma(\bm{\xi})\right)^\top& = \bm{\eta}_i^\top\Sigma^{-1}X_i.
\end{align}

\subsection{Computing $\nabla_{\xi}\log \phi_{J}\left(\bm{z}_i;X_i\bm{\beta},\Sigma(\bm{\xi})\right)$}
We use the convention that for two generic matrices $C_{\text{dim1}\times\text{dim2}}$ and $L_{\text{dim3}\times\text{dim4}}$ we have that
$$\frac{\partial C}{\partial L} = \frac{\partial \text{vec}(C)}{\partial \text{vec}(L)} = E,$$
where $E$ is of dimension $(\text{dim1}*\text{dim2})\times(\text{dim3}*\text{dim4})$.
We apply the chain rule of derivatives multiple times to write 
\begin{align}
    \nabla_{\xi}\log \phi_{J}\left(\bm{z}_i;X_i\bm{\beta},\Sigma(\bm{\xi})\right)^\top &= \frac{\partial}{\partial \bm{\xi}}\left[-\frac{J}{2}\log(2\pi)-\frac{1}{2}\log (\text{det}(\Sigma))-\frac{1}{2}\bm{\eta}_i^\top\Sigma^{-1}\bm{\eta}_i\right]\nonumber\\
    & =  -\frac{1}{2}\frac{\partial}{\partial \bm{\xi}}\log (\text{det}(\Sigma))-\frac{1}{2}\frac{\partial}{\partial \bm{\xi}}\bm{\eta}_i^\top\Sigma^{-1}\bm{\eta}_i\nonumber\\
    & =  -\frac{1}{2}\frac{1}{\text{det}(\Sigma)}\frac{\partial \text{det}(\Sigma)}{\partial \Sigma}\frac{\partial \Sigma}{\partial \bm{\xi}}-\frac{1}{2}\frac{\partial}{\partial \bm{\xi}}\bm{\eta}_i^\top\Sigma^{-1}\bm{\eta}_i\nonumber \\
    & =  -\frac{1}{2}\frac{1}{\text{det}(\Sigma)}\frac{\partial \text{det}(\Sigma)}{\partial \Sigma}\frac{\partial \Sigma}{\partial \bm{\xi}}-\frac{1}{2}\left(\bm{\eta}_i^\top\otimes\bm{\eta}_i^\top\right)\frac{\partial}{\partial \bm{\xi}}\text{vec}(\Sigma^{-1})\nonumber\\
     & =  -\frac{1}{2}\frac{1}{\text{det}(\Sigma)}\frac{\partial \text{det}(\Sigma)}{\partial \Sigma}\frac{\partial \Sigma}{\partial \bm{\xi}}-\frac{1}{2}\left(\bm{\eta}_i^\top\otimes\bm{\eta}_i^\top\right)\frac{\partial \Sigma^{-1}}{\partial \Sigma}\frac{\partial \Sigma}{\partial \bm{\xi}}\nonumber\\
     & =  -\frac{1}{2}\frac{1}{\text{det}(\Sigma)}\frac{\partial \text{det}(\Sigma)}{\partial \Sigma}\frac{\partial \Sigma}{\partial B}\frac{\partial B}{\partial\bm{\kappa}}\frac{\partial \bm{\kappa}}{\partial \bm{\xi}}-\frac{1}{2}\frac{1}{\text{det}(\Sigma)}\frac{\partial \text{det}(\Sigma)}{\partial \Sigma}\frac{\partial \Sigma}{\partial \bm{d}}\frac{\partial \bm{d}}{\partial\bm{\kappa}}\frac{\partial \bm{\kappa}}{\partial \bm{\xi}}\nonumber\\
     &\hspace{0.7cm}-\frac{1}{2}\left(\bm{\eta}_i^\top\otimes\bm{\eta}_i^\top\right)\frac{\partial \Sigma^{-1}}{\partial \Sigma}\frac{\partial \Sigma}{\partial B}\frac{\partial B}{\partial\bm{\kappa}}\frac{\partial \bm{\kappa}}{\partial \bm{\xi}}-\frac{1}{2}\left(\bm{\eta}_i^\top\otimes\bm{\eta}_i^\top\right)\frac{\partial \Sigma^{-1}}{\partial \Sigma}\frac{\partial \Sigma}{\partial \bm{d}}\frac{\partial \bm{d}}{\partial\bm{\kappa}}\frac{\partial \bm{\kappa}}{\partial \bm{\xi}}\nonumber,
\end{align}
where the fourth line uses that vec($AXB^\top)=(B\otimes A)\text{vec}(X)$.
The sixth line uses that $\frac{\partial \Sigma}{\partial \bm{\xi}} = \frac{\partial \Sigma}{\partial B}\frac{\partial B}{\partial \bm{\kappa}}\frac{\partial \bm{\kappa}}{\partial \bm{\xi}}+\frac{\partial \Sigma}{\partial \bm{d}}\frac{\partial \bm{d}}{\partial \bm{\kappa}}\frac{\partial \bm{\kappa}}{\partial \bm{\xi}}$. Note that $\frac{\partial \text{det}(\Sigma)}{\partial \Sigma} = \text{det}(\Sigma)\text{vec}(\Sigma^{-1})^\top$. We derive the remaining expressions below. 

\subsubsection{Derivation of  $\frac{1}{2}\left(\bm{\eta}_i^\top\otimes\bm{\eta}_i^\top\right)\frac{\partial \Sigma^{-1}}{\partial \Sigma}\frac{\partial \Sigma}{\partial B}$}

We have
$\frac{\partial \Sigma^{-1}}{\partial \Sigma} = -\left(\Sigma^{-1}\otimes\Sigma^{-1}\right)$ and
$\frac{\partial \Sigma}{\partial B} = \left(I_{J^2}+K_{J,J}\right)\left(B\otimes I_J\right)$,
where $K_{m,n}$ is the commutation matrix of an $m\times n$ matrix.
The product of these two derivatives can be simplified as
\begin{align}
\frac{\partial \Sigma^{-1}}{\partial \Sigma}\frac{\partial \Sigma}{\partial B} = -\left(I_{J^2}+K_{J,J}\right)\left(\Sigma^{-1}B\otimes\Sigma^{-1}\right).
\end{align}
It follows that
\begin{align}
    \frac{1}{2}\left(\bm{\eta}_i^\top\otimes\bm{\eta}_i^\top\right)\frac{\partial \Sigma^{-1}}{\partial \Sigma}\frac{\partial \Sigma}{\partial B}&=-\frac{1}{2}\left(\bm{\eta}_i^\top\otimes\bm{\eta}_i^\top\right)\left(I_{J^2}+K_{J,J}\right)\left(\Sigma^{-1}B\otimes\Sigma^{-1}\right)\\
    & =-\frac{1}{2}\left(\bm{\eta}_i^\top\Sigma^{-1}B\otimes\bm{\eta}_i^\top\Sigma^{-1}\right)-\frac{1}{2}\left(\bm{\eta}_i^\top\Sigma^{-1}\otimes\bm{\eta}_i^\top\Sigma^{-1}B\right)K_{J,q}\\
    & = -\left(\bm{\eta}_i^\top\Sigma^{-1}B\otimes\bm{\eta}_i^\top\Sigma^{-1}\right).
\end{align}

\subsubsection{Derivation of  $\frac{1}{2}\left(\bm{\eta}_i^\top\otimes\bm{\eta}_i^\top\right)\frac{\partial \Sigma^{-1}}{\partial \Sigma}\frac{\partial \Sigma}{\partial \bm{d}}$}
Following identical steps as above we can show that 
\begin{equation}
    \frac{1}{2}\left(\bm{\eta}_i^\top\otimes\bm{\eta}_i^\top\right)\frac{\partial \Sigma^{-1}}{\partial \bm{d}}
    =\frac{1}{2}\left(\bm{\eta}_i^\top\otimes\bm{\eta}_i^\top\right)\frac{\partial \Sigma^{-1}}{\partial D} P= -\left(\bm{\eta}_i^\top\Sigma^{-1}D\otimes\bm{\eta}_i^\top\Sigma^{-1}\right)P,
\end{equation}
where $P$ extracts the columns corresponding to the elements of $\bm{d}$.

\subsubsection{Derivation of $\frac{\partial B}{\partial \bm{\kappa}}$}
Denote $C_k =  \left[B_k |\bm{d}_k\right]$ and define $C = \left[C_1^\top |\dots| C_K^\top\right]^\top$. Note that $B = CP_3$, $\bm{d} =CP_4$ for $P_3 = [I_{q}|\bm{0}_{q\times 1}]^\top$ and $P_2 = \left(\bm{0}_{1\times q},1\right)^\top$. With this notation we can then write
\begin{align}
\frac{\partial B}{\partial \bm{\kappa}}=\frac{\partial B}{\partial C}\frac{\partial C}{\partial \bm{\kappa}}.    
\end{align}
The first term can be derived from the expression
\begin{align}
    \text{vec}\left(B\right)&=\left(P_3^\top\otimes I_{J}\right)\text{vec}\left(C\right),
\end{align}
from which we obtain 
$\frac{\partial B}{\partial C}=\left(P_3^\top\otimes I_{J}\right)$.
The second term can be derived as 
$$\frac{\partial C}{\partial \bm{\kappa}} = K_{q+1,J}\frac{\partial C^\top}{\partial \bm{\kappa}}$$
where
\begin{align}
\frac{\partial C^\top}{\partial \bm{\kappa}} = \text{blockdiag}\left(\frac{\partial C_1^\top}{\partial \bm{\kappa}_1},\dots,\frac{\partial C_K^\top}{\partial \bm{\kappa}_K}\right),
\end{align}
$\frac{\partial C_k^\top}{\partial \bm{\kappa}_k} =K_{J_k,q+1} \frac{\partial C_k}{\partial \bm{\kappa}_k}$, and the elements of $\frac{\partial C_k}{\partial \bm{\kappa}_k}$ equal 
$\left\{\frac{\partial C_k}{\partial \bm{\kappa}_k}\right\}_{l,j}=\frac{\partial \psi_{kl}(\bm{\kappa}_k)}{\partial \kappa_{k,j}}$ with
\begin{align}
\frac{\partial \psi_{kl}(\bm{\kappa}_k)}{\partial \kappa_{kj}}=
\left\{
\begin{array}{ll}
\sqrt{J_k}\cos\left(\kappa_{kj}\right)\cos\left(\kappa_{kl}\right)\prod_{s\in\{1,\dots,l-1\}\backslash j}\sin\left(\kappa_{ks}\right) & \text{if } j<l \text{ and } l<n_k,\\
-\sqrt{J_k}\prod_{s=1}^{l}\sin\left(\kappa_{ks}\right) & \text{if } j=l \text{ and } l<n_k,\\
\sqrt{J_k}\cos\left(\kappa_{kj}\right)\prod_{s\in\{1,\dots,l-1\}\backslash j}\sin\left(\kappa_{ks}\right) & \text{if } j<l \text{ and } l=n_k,\\
0 & \text{if otherwise.} 
\end{array}
\right.
\end{align}


\subsubsection{Derivation of $\frac{\partial \bm{d}}{\partial \bm{\kappa}}$}
Similarily, we have that
\begin{align}
    \frac{\partial \bm{d}}{\partial \bm{\kappa}}=\frac{\partial \bm{d}}{\partial C}\frac{\partial C}{\partial \bm{\kappa}}.
\end{align}
We calculated the second term previously. Following the same logic as before, the first term can be computed from noting that
\begin{align}
\text{vec}\left(\bm{d}\right)=\left(P_4^\top\otimes I_J\right)\text{vec}\left(C\right),
\end{align}
so that 
$\frac{\partial \bm{d}}{\partial C}=\left(P_4^\top\otimes I_J\right)$.




\section{VB with the identity covariance matrix}\label{A:VI_I}
The variational approach proposed in this paper can also be applied to estimation of the MVMNP model with identity covariance matrix. The main difference is that this model does not have a vector of angles $\bm{\kappa}$. 
When $\Sigma$ is fixed at the identity matrix, VB only requires an unbiased estimate of the gradient ${\nabla_{\beta}\log g}\left(\bm{\beta},\bm{z}\right)$. The function $\log g(\bm{\beta},\bm{z})^\top$ can be written as 
\begin{equation}
 \log g(\bm{\beta},\bm{z}) = \log p(\bm{y}|\bm{z})+ \log p(\bm{\beta})+\sum_{i=1}^{N}\log \phi_{J}\left(\bm{z}_i;X_i\bm{\beta},I_J\right).
\end{equation}

The estimate can be constructed on a set $A\subset\{1,\dots,N\}$ of $M$ indexes sampled at random and without replacement:
\begin{align}
    \widehat{\nabla_{\beta}\log g(\bm{\beta},\bm{z}_A)}& = \nabla_{\beta}\log p(\bm{\beta})+\frac{N}{M}\sum_{i\in A}\nabla_{\beta}\log \phi_{J}\left(\bm{z}_i;X_i\bm{\beta},I_J\right),
\end{align}

Appendix~\ref{A:dbetaprior} shows how to evaluate $\nabla_{\beta}\log p(\bm{\beta})$, and Appendix~\ref{A:dbetaZ} provides an expression for $\nabla_{\beta}\log \phi_{J}\left(\bm{z}_i;X_i\bm{\beta},\Sigma(\bm{\xi})\right)$, which after replacing  $\Sigma(\bm{\xi})=I_J$ provides the required gradient. 

\section{Additional results numerical experiments}\label{A:simulation}
\begin{figure}[h!]
\caption{Posterior means and standard deviations in numerical experiment ($N = 10,000$)}
\centering
\includegraphics*[width=\textwidth,trim = 0 0 0 0]{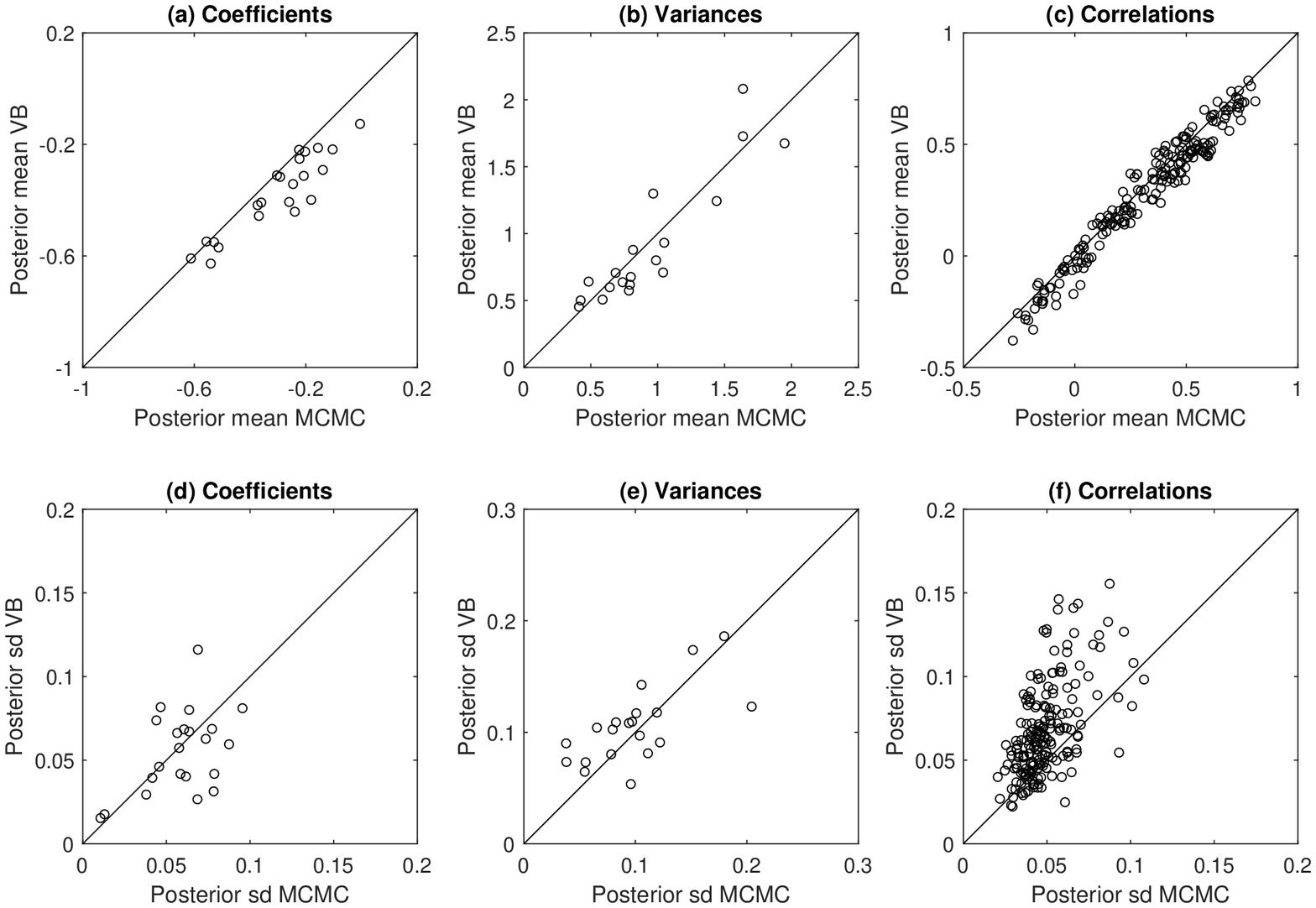}
 \fnote{Panels (a) to (c) present the estimated posterior means from MCMC (x-axis) against those from VB(10\%) (y-axis). Panels (d) to (f) show corresponding plots for the posterior standard deviations. Panels (a) and (d) correspond to $\bm\beta$, Panels (b) and (e) correspond to the diagonal elements of $\Sigma$, and Panels (c) and (f) to the implied correlations.
 }
\label{fig:param1}
\end{figure}

\begin{figure}[h!]
\caption{Posterior means and standard deviations in numerical experiment ($N = 10,000$)}
\centering
\includegraphics*[width=\textwidth,trim = 0 0 0 0]{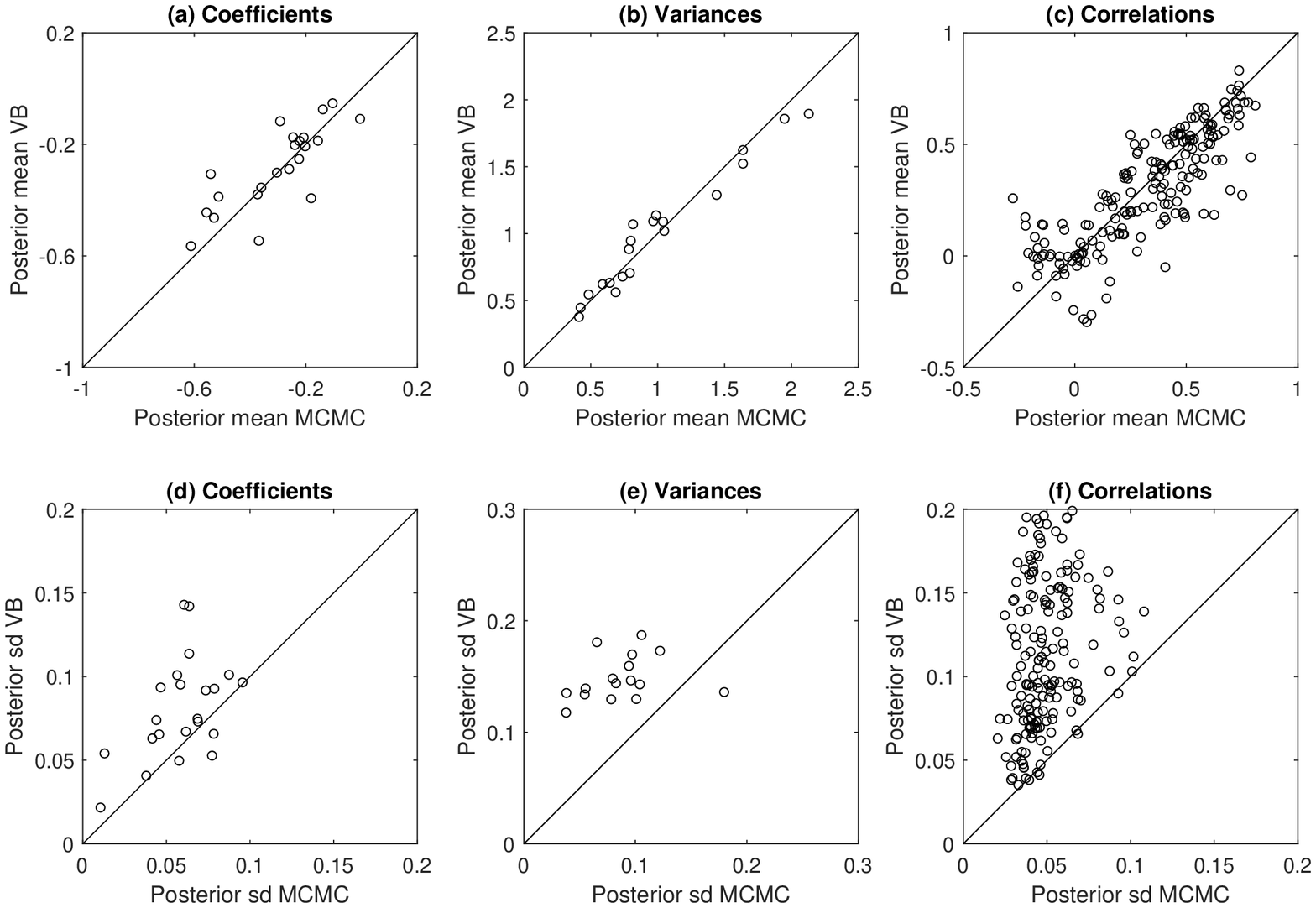}
 \fnote{Panels (a) to (c) present the estimated posterior means from MCMC (x-axis) against those from VB(1\%) (y-axis). Panels (d) to (f) show corresponding plots for the posterior standard deviations. Panels (a) and (d) correspond to $\bm\beta$, Panels (b) and (e) correspond to the diagonal elements of $\Sigma$, and Panels (c) and (f) to the implied correlations.
 }
\label{fig:param2}
\end{figure}

\begin{figure}[h!]
\caption{Sensitivity of out-of-sample log-score to the choice of the number of factors in simulation exercise}
\centering
\includegraphics*[width=\textwidth,trim = 0 0 0 0]{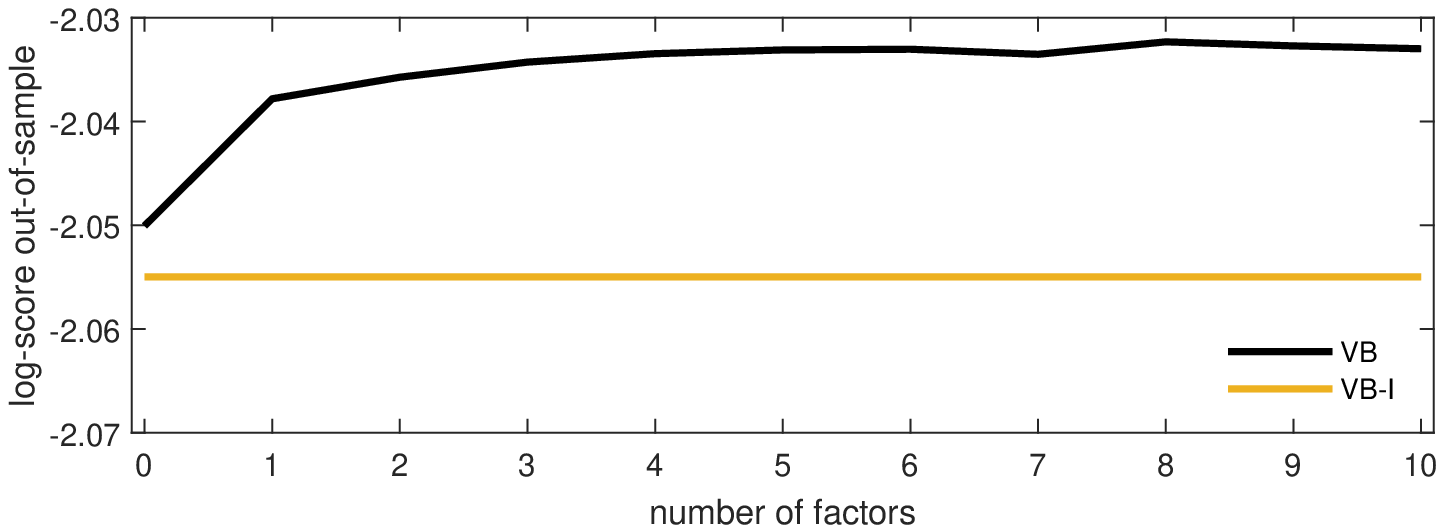}
 \fnote{The black line presents the out-of-sample log-score averaged over choices 1 and 2, against the total number of factors $p$ for VB. The yellow horizontal line shows this measure for VB-I as a benchmark.}
\label{fig:ls_robustness}
\end{figure}

\clearpage
\section{Additional results empirical applications}\label{A:empirical}


\begin{figure}[h!]
\caption{Posterior mean parameters in laundry detergent application}
\centering
\includegraphics*[width=\textwidth,trim = 0 0 0 0]{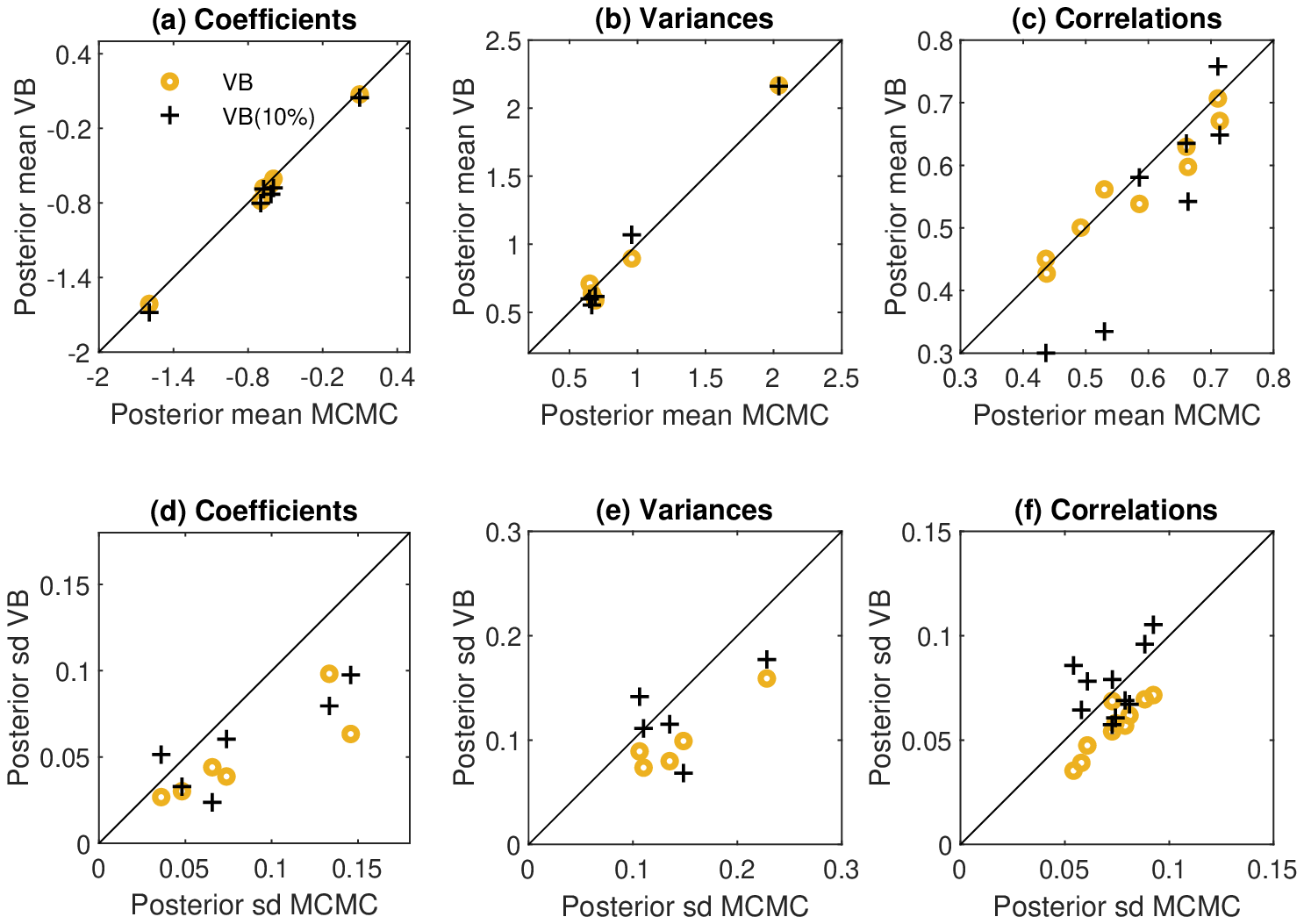}
 \fnote{This figure presents the estimated posterior means from VB (yellow circles) and VB(10\%) (black crosses) on the y-axis, and the estimated posterior means from MCMC on the x-axis, for the coefficients $\beta$ in Panel (a), and the variances and correlations of the latent utilities in $\Sigma$ in Panel (b) and (c), respectively. Panels (d) to (f) present the corresponding results for the posterior standard deviations.}
\label{fig:param_app_small}
\end{figure}

\begin{figure}[tb!]
\caption{Posterior mean parameters in pasta application}
\centering
\includegraphics*[width=\textwidth,trim = 0 0 0 0]{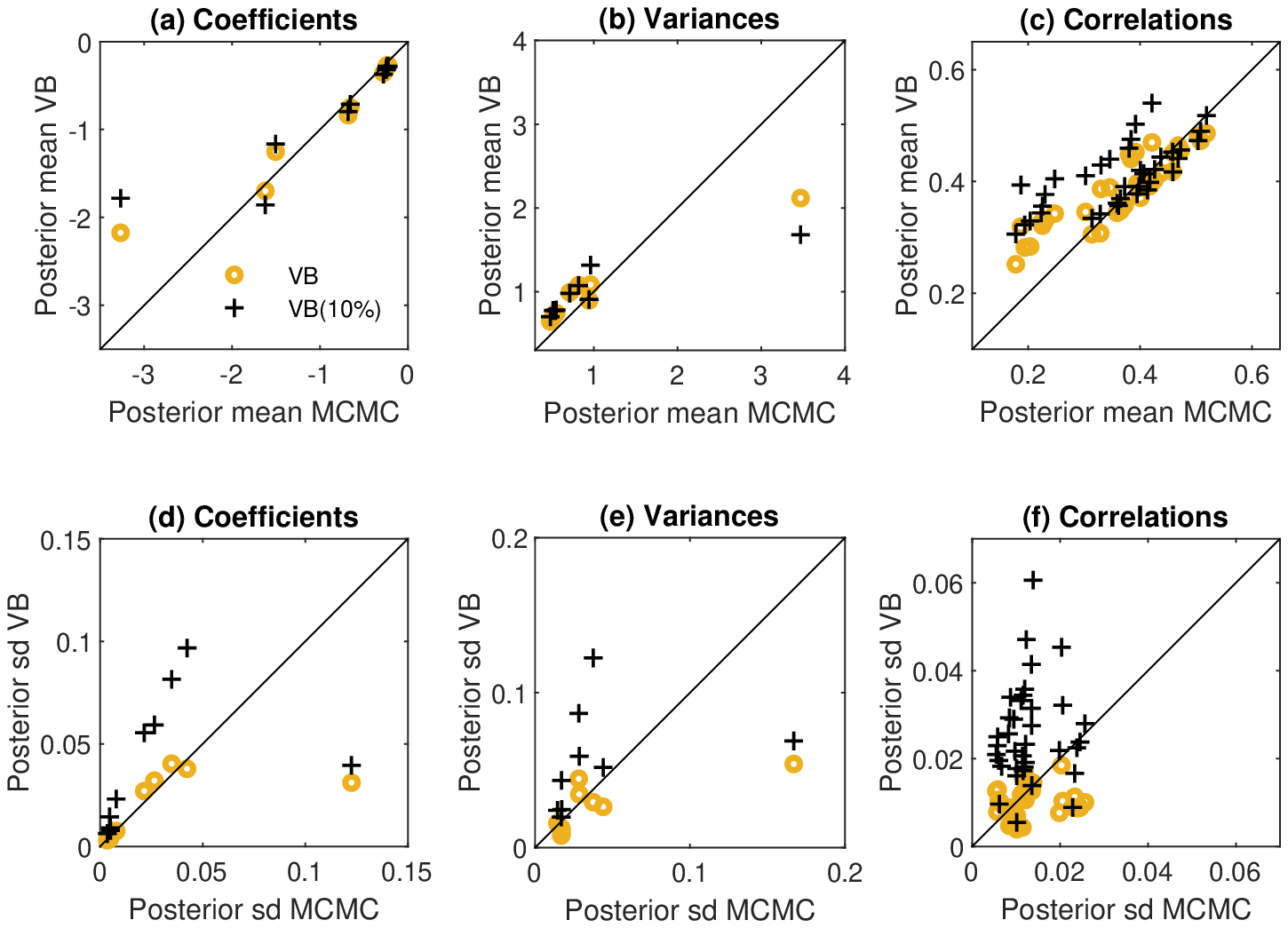}
 \fnote{This figure presents the estimated posterior means from VB(10\%) (yellow circles) and VB(1\%) (black crosses) on the y-axis, and the estimated posterior means from MCMC on the x-axis, for the coefficients $\beta$ in Panel (a), and the variances and correlations of the latent utilities in $\Sigma$ in Panel (b) and (c), respectively.
 }
\label{fig:param_app_big}
\end{figure}




\end{document}